\documentclass[twocolumn,twocolappendix]{aastex631}

\usepackage{amsmath}
\usepackage{bm}
\usepackage{graphicx}

\newcommand{\cs}{c_{\rm s}}

\newcommand{\mbh}{M_\bullet}
\newcommand{\msun}{M_\odot}
\newcommand{\vinf}{v_\infty}

\begin{document}


\title{Formation of black hole stars via star--black hole collisions}

\correspondingauthor{Yanlong Shi}
\email{yanlong@cita.utoronto.ca}

\newcommand{\cita}{Canadian Institute for Theoretical Astrophysics, University of Toronto, Toronto, ON M5S 3H8, Canada}
\newcommand{\caltech}{TAPIR, MC 350-17, California Institute of Technology, Pasadena, CA 91125, USA}

\newcommand{\ucsc}{Department of Astronomy and Astrophysics, University of California, Santa Cruz, CA, 95064, USA}

\newcommand{\tsinghuaias}{Institute for Advanced Study, Tsinghua University, Beijing, 100084, China}

\newcommand{\tsinghuadoa}{Department of Astronomy, Tsinghua University, Beijing, 100084, China}

\newcommand{\westlake}{Department of Astronomy, Westlake University, Hangzhou, Zhejiang, 310030, China}

\newcommand{\ucsb}{Department of Physics, University of California, Santa Barbara, CA 93106, USA}

\author[0000-0002-0087-3237]{Yanlong Shi}
\affiliation{\cita}

\author[0009-0007-9015-9451]{Qingru Hu}
\affiliation{\tsinghuadoa}

\author{Zhenghao Xu}
\affiliation{\ucsb}
\affiliation{\tsinghuadoa}

\author[0000-0001-5466-4628]{Douglas N. C. Lin}
\affiliation{\westlake}
\affiliation{\ucsc}
\affiliation{\tsinghuaias}

\author[0000-0002-8659-3729]{Norman Murray}
\affiliation{\cita}





\begin{abstract}

In dense stellar environments such as globular clusters and active galactic nucleus (AGN) disks, stellar-mass black holes (sBHs) may frequently collide with massive stars. We investigate this process using semi-analytic models, three-dimensional hydrodynamical simulations, and one-dimensional stellar evolution calculations, focusing on collisions between sBHs and a $M_\star=100\,M_\odot$ main-sequence star with a radius $R_\star=16\,R_\odot$. We find that gas drag retains the BH within the stellar envelope unless the impact velocity exceeds $\sim2\sqrt{G(M_\star+M_\bullet)/R_\star}$. The post-collision outcome depends primarily on the BH-to-star mass ratio. For $M_\bullet\gtrsim30\,M_\odot$, the retained envelope is either quasi-spherical or disk-like, but remains dynamically unstable because of shock heating. In contrast, for $M_\bullet\lesssim10\,M_\odot$, the collision forms a ``black hole star'' (BH*): a quasi-hydrostatic, extended stellar envelope surrounding the embedded BH. These results agree with our analytic prediction that BH* formation necessarily requires $M_\bullet\lesssim0.2\,M_\star$. Follow-up \texttt{MESA} calculations further show that, for these low-mass BHs, the shock-heated remnant thermally relaxes without triggering runaway expansion. We discuss several astrophysical implications of BH*s, including their evolution, the possibility of gravitational-wave events from BH binaries assembled within a stellar envelope, and repeated star--sBH collisions as a pathway for rapid BH growth in dense stellar systems. This mechanism may contribute to the formation of massive BHs in high-redshift nuclear star clusters and may be relevant to the origin of the ``little red dots'' discovered by JWST.
\end{abstract}



\section{Introduction} 
\label{sec:intro}

The recent discovery by JWST of a population of compact, red sources at high redshift ($z \gtrsim 4$), dubbed ``little red dots'' \citep[LRDs, e.g.,][]{MattheeNaiduBrammer_2024ApJ...963..129M,KokorevCaputiGreene_2024ApJ...968...38K,KocevskiFinkelsteinBarro_2025ApJ...986..126K}, has raised fundamental questions about the evolution of black holes (BHs) in the early Universe. Apart from their compact morphology, LRDs exhibit unique spectral features, like a blue rest-frame UV continuum declining with frequency, a red optical slope with a prominent Balmer break, and broad hydrogen emission lines indicative of gas velocities $\sim 1000\,\rm km\,s^{-1}$. Among many possible scenarios of LRDs \citep[][]{InayoshiHo_2025arXiv251203130I}, one compelling resolution is the ``black hole star'' (or ``quasi-star,'' we abbreviate it as ``BH*'' thereafter) interpretation \citep[e.g.,][]{deGraaffHvidingNaidu_2026MNRAS.551g1567D,BegelmanDexter_2026ApJ...996...48B,InayoshiMuraseKashiyama_2026ApJ..1000...90I,SunNaiduMatthee_2026OJAp....962505S,NaiduMattheeKatz_2026Natur.656..329N}: an intermediate-mass/supermassive black hole (IMBH/SMBH) embedded in a massive gaseous envelope simultaneously produces some of the key features: e.g., the gaseous envelope sets $T_{\rm eff} \sim 5000\,\rm K$ blackbody radiation and reproduces the observed red continuum \citep[e.g.,][]{BegelmanDexter_2026ApJ...996...48B,InayoshiMuraseKashiyama_2026ApJ..1000...90I,deGraaffHvidingNaidu_2026MNRAS.551g1567D,YanagisawaOuchiGolubchik_2026arXiv260106015Y}; the extended stellar atmosphere gives rise to the Balmer break through hydrogen bound-free absorption \citep[][]{InayoshiMaiolino_2025ApJ...980L..27I,NaiduMattheeKatz_2026Natur.656..329N,deGraaffRixNaidu_2025A&A...701A.168D,SunNaiduMatthee_2026OJAp....962505S}; the virial motion and electron scattering outside the envelope explain the broad emission lines \citep[][]{InayoshiMuraseKashiyama_2026ApJ..1000...90I,KokorevChisholmNaidu_2026ApJ..1004..153K}. 

BH*s are not merely an ad hoc solution to the LRD puzzle, since the theoretical possibility of compact objects embedded in stellar envelopes has a long history. The concept was pioneered by \citet{ThorneZytkow_1977ApJ...212..832T}, who showed that a neutron star engulfed by a giant stellar envelope can reach a quasi-hydrostatic equilibrium, forming a Thorne--\.Zytkow object \citep[T\.ZO;][]{ThorneZytkow_1977ApJ...212..832T,HiraiPodsiadlowski_2022MNRAS.517.4544H}. T\.ZOs produce distinctive nucleosynthetic signatures that have been tentatively identified in observations \citep[e.g.,][]{LevesqueMasseyZytkow_2014MNRAS.443L..94L}. BH*s are the natural higher-mass analogue: rather than a neutron star, a BH sits at the center of the envelope, with its accretion luminosity providing additional pressure support in place of nuclear shell burning \citep[e.g.,][]{BegelmanRossiArmitage_2008MNRAS.387.1649B,VolonteriBegelman_2010MNRAS.409.1022V,BallToutZytkow_2011MNRAS.414.2751B,CoughlinBegelman_2024ApJ...970..158C,HassanPernaCantiello_2026ApJ...998...65H}. 

A natural formation channel for BH*s is the direct collision between a stellar-mass BH (sBH) and a massive star \citep[e.g.,][]{Rantala_2026arXiv260422924R}. Such collisions are expected to be frequent in two broad classes of dense stellar environment: nuclear star clusters and globular clusters, where high stellar densities and velocity dispersions drive repeated dynamical encounters \citep{BinneyTremaine_2008gady.book.....B,Rantala_2026arXiv260422924R}, and AGN disks, where the gas-rich environment traps stars 
and compact objects in a common plane and further enhances collision rates \citep[e.g.,][]{TagawaHaimanKocsis_2020ApJ...898...25T,YangBartosFragione_2022ApJ...933L..28Y,ChenLin_2024ApJ...967...88C}. Depending on the BH mass, impact parameter, and relative velocity, a star--sBH collision can result in one of several outcomes \citep[refer to discussions in][]{Rantala_2026arXiv260422924R}. (1)~When both objects closely interact but not yet geometrically collide, the star may undergo a ``micro tidal disruption event'' \citep[micro-TDE; e.g.,][]{PeretsLiLombardi_2016ApJ...823..113P,KremerLuRodriguez_2019ApJ...881...75K,RyuPernaWang_2022MNRAS.516.2204R,VynatheyaRyuPakmor_2024A&A...685A..45V,XinHaimanPerna_2024ApJ...961..149X,KirogluKremerBiscoveanu_2025ApJ...979..237K,RastelloIorioGieles_2026A&A...707A.217R} in the most violent limit. (2)~When the BH and the star geometrically collide, a more violent disruption of the star may happen \citep[e.g.,][]{KremerLombardiLu_2022ApJ...933..203K,HiraiPodsiadlowski_2022MNRAS.517.4544H,TsunaLu_2025ApJ...986...84T}; repeated star--sBH collisions may boost the formation of IMBHs \citep[][]{RoseNaozSari_2022ApJ...929L..22R}. (3)~On the other hand, in less energetic collisions (the released energy is insufficient to disrupt the star), gas drag may decelerate the BH sufficiently for it to be retained within the stellar envelope, forming a BH*---which we focus on in this paper.

The gas drag that governs BH retention is physically related to the drag force studied extensively in the context of common-envelope (CE) evolution \citep[e.g.,][]{IvanovaJusthamChen_2013A&ARv..21...59I}, in which a BH companion spirals inward through the envelope of a giant star. In CE studies, the drag is typically modeled in the Bondi--Hoyle--Lyttleton \citep[BHL;][]{HoyleLyttleton_1939PCPS...35..405H,Bondi_1952MNRAS.112..195B,Edgar_2004NewAR..48..843E} framework or via the energy formalism, which sets the inspiral timescale and envelope ejection \citep[e.g.,][]{RickerTaam_2012ApJ...746...74R,MacLeodRamirez-Ruiz_2015ApJ...803...41M,Cruz-OsorioRezzolla_2020ApJ...894..147C}. The star--sBH collision problem shares this drag physics but operates in a qualitatively different regime, since the BH enters the star impulsively at high velocity, depositing energy through a strong shock on dynamical timescales. 

Prior simulation works have studied star--sBH collisions primarily in the tidal disruption and electromagnetic transient regime \citep[e.g.,][]{KremerLombardiLu_2022ApJ...933..203K,VynatheyaRyuPakmor_2024A&A...685A..45V,XinHaimanPerna_2024ApJ...961..149X}, while the conditions for BH retention and the resulting post-collision structure have not been systematically explored. In this paper, we address these questions through a combined semi-analytic and hydrodynamical study of collisions between sBHs of mass $5$--$100\,M_\odot$ and a $100\,M_\odot$ main-sequence star, a representative mass for massive stars in both AGN disks and nuclear star clusters, where runaway stellar mergers preferentially build up the most massive objects \citep[e.g.,][]{PortegiesZwartMcMillan_2002ApJ...576..899P,KremerSperaBecker_2020ApJ...903...45K,ShiGrudicHopkins_2021MNRAS.505.2753S,RantalaNaabLahen_2024MNRAS.531.3770R}. Our semi-analytic approach integrates the BH equation of motion under BHL drag, enabling an efficient survey of the full parameter space of BH mass, impact parameter, and initial velocity. We complement this with three-dimensional hydrodynamical simulations using the \texttt{GIZMO} code \citep[]{Hopkins_2015MNRAS.450...53H}, which resolve the BH dynamics, shock structure, and envelope morphology for a representative subset of cases.

The remainder of this paper is organized as follows. Section~\ref{sec:semi-analytic} presents the semi-analytic model and stellar evolution calculations, focusing on the orbital dynamics and energy budget of the collision. Section~\ref{sec:hydro} describes the hydrodynamical simulations. Section~\ref{sec:implications} discusses the astrophysical implications of such star--sBH collisions. Section~\ref{sec:conclusions} summarizes our findings and future directions.

\begin{figure*}
    \includegraphics[width=\linewidth]{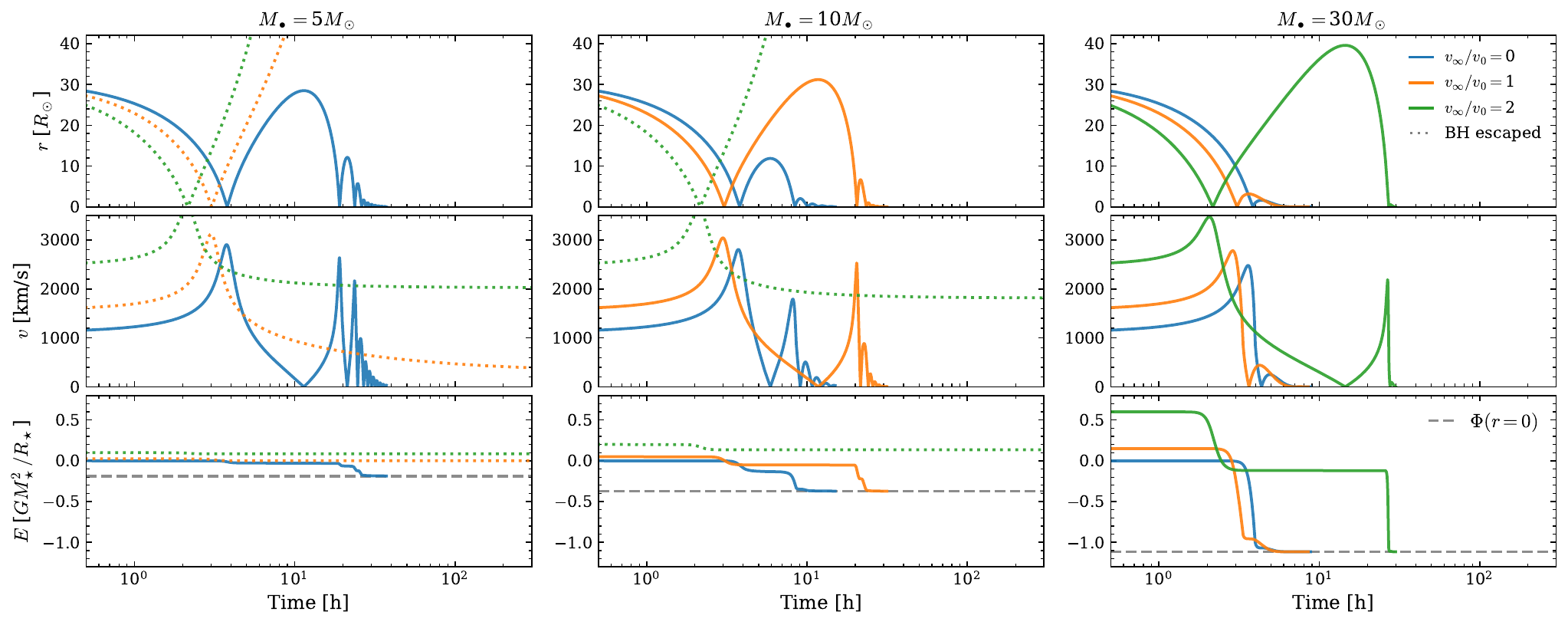}
    \caption{Position (\emph{top panels}), velocity (\emph{middle panels}) and energy (\emph{bottom panels}) evolution for head-on ($b= 0$) star--sBH collisions. Blue, orange, and green lines correspond to $v_\infty / v_0 = 0$, $1$, and $2$, respectively. Dotted lines indicate semi-analytic simulations where the sBH escapes the star. The gray dashed line in the bottom panel marks the minimum potential at the stellar center $\Phi(r=0)$.}
    \label{fig:b-0.0}
\end{figure*}

\section{Analytical Expectations}
\label{sec:semi-analytic}

\subsection{The gas dynamical friction}\label{sect:BHL-drag}
As an sBH moves through the gas envelope of a massive star, it gravitationally focuses the gas flow into an overdense wake behind it. The wake exerts a gravitational drag on the moving sBH, known as dynamical friction \citep{Chandrasekhar_1943ApJ....97..255C,Ostriker_1999ApJ...513..252O}.
In the highly supersonic regime, where gas pressure is negligible, \citet{HoyleLyttleton_1939PCPS...35..405H} estimated the cross section of this gravitational focusing with the Hoyle--Lyttleton (HL) radius $R_{\rm HL}=2G\mbh/v^2$, where $\mbh$ is the sBH mass and $v$ its speed relative to the gas, yielding a drag force $F_{\rm HL}\sim\pi R_{\rm HL}^2\rho v^2 = 4\pi G^2\mbh^2\rho/v^2$, with $\rho$ the gas density at the sBH location. While this estimate diverges as $v\to 0$, at mild Mach numbers, the work of \citet{Bondi_1952MNRAS.112..195B} leads to the interpolation formula
\begin{equation}\label{equ:drag-force}
    F_{\rm drag} \sim \frac{4\pi G^2 \mbh^2 \rho\, v}{(\cs^2+v^2)^{3/2}},
\end{equation}
where $\cs$ is the gas sound speed, recovering $F_{\rm HL}$ when $v\gg\cs$ \citep{Edgar_2004NewAR..48..843E}. This estimate of the drag forces sets the baseline for a series of succeeding works of linear
perturbation analysis \citep{Ostriker_1999ApJ...513..252O} and numerical simulations \citep[e.g.,][]{LescaudronDuboisBeckmann_2023A&A...674A.217L,SuzuguchiSugimuraHosokawa_2024ApJ...966....7S}, as also reviewed by \citet{RopkeDeMarco_2023LRCA....9....2R}.

The drag force dissipates the sBH's orbital energy at a rate $\dot E_{\rm drag} \sim F_{\rm drag} v$, also referred to as the dynamical friction luminosity, so the sBH kinetic energy is damped on a timescale of
\begin{equation}
    \tau_{\rm damp} \sim \frac{\mbh v^2/2}{\dot E_{\rm drag}}
    = \frac{(\cs^2+v^2)^{3/2}}{8\pi G^2 \rho \mbh}.\label{equ:stoping_time}
\end{equation}
For typical conditions inside a massive star, e.g., $\rho = 0.2\,{\rm g\,cm^{-3}}$ and $\cs = 500\,{\rm km\,s^{-1}}$, an sBH with $\mbh = 5\,M_\odot$ moving transonically ($v \approx \cs$) is stopped within $\tau_{\rm damp} \approx 0.5\,{\rm h}$, comparable to or shorter than the stellar dynamical time. Such efficient dissipation implies that gas dynamical friction alone may fully damp the orbital energy of an sBH colliding with a massive star, retaining it inside the star. In the following part, we explicitly calculate the orbits of the sBH colliding with a massive star to further quantify this process.

\subsection{Semi-analytic orbital integration}\label{subsect:orb-integ}

\subsubsection{Model setups and initial condition}
\label{sect:initial-condidtions}

We consider the simplified limit $\mbh \ll M_\star$, in which the massive star
can be treated as a fixed background gas cloud when integrating the orbital
trajectory of the sBH. This semi-analytic approach is similar to that utilized in CE studies \citep[e.g.,][]{MacLeodRamirez-Ruiz_2015ApJ...803...41M,MacLeodRamirez-Ruiz_2015ApJ...798L..19M}. The sBH
then evolves under the gravity of the background star and the gaseous drag
force,
\begin{equation}\label{equ:eom}
    \ddot{\bm{r}} = -\frac{G M_\star(<r)}{r^3}\,\bm{r}
    + \frac{\bm{F}_{\rm drag}}{\mbh},
\end{equation}
where $\bm{r}$ is the sBH position relative to the stellar center and
$M_\star(<r)$ is the enclosed stellar mass within radius $r$. Following
\citet{Ostriker_1999ApJ...513..252O}, the drag force is
\begin{equation}
    \bm{F}_{\rm drag} = -\frac{4\pi G^2 \mbh^2 \rho}{v^3}\,
    I(\mathcal{M})\,\bm{v},
\end{equation}
where $\bm{v}$ is the sBH velocity, and the prefactor $I(\mathcal{M})$ encodes the dependence on
the Mach number $\mathcal{M} \equiv v/\cs$. Since the original
$I(\mathcal{M})$ of \citet{Ostriker_1999ApJ...513..252O} is singular at
$\mathcal{M} = 1$, we adopt the regularized form of \citet{GenerozovPerets_2023MNRAS.522.1763G},
\begin{equation}
I(\mathcal{M}) =
\begin{cases}
\ln\Lambda, & \mathcal{M} \geq 1; \\[4pt]
\min\left[\ln\Lambda,\
\dfrac{1}{2}\ln\left(\dfrac{1+\mathcal{M}}{1-\mathcal{M}}\right)
- \mathcal{M}\right], & \mathcal{M} < 1,
\end{cases}
\end{equation}
where the Coulomb logarithm is set to
$\ln\Lambda = 1.4 \sim \ln(R_\star/R_{\rm HL})$, the ratio of the largest
to the smallest radial scales contributing to the drag.

We integrate Equation~\eqref{equ:eom} for the sBH orbit using the
\texttt{REBOUNDx} package \citep{TamayoReinShi_2020MNRAS.491.2885T}, an extension of the
\texttt{REBOUND} code \citep{ReinLiu_2012A&A...537A.128R}, with the 15th-order \texttt{IAS15} integrator \citep{ReinSpiegel_2015MNRAS.446.1424R}. The integration terminates when either (1) the distance between the sBH and the stellar center falls below $10^{-3}\,R_\odot$, indicating that the sBH has settled at the stellar center, or (2) the integration time reaches $10^3$ hours. We mark the time when the simulation terminates as the ``stopping time'' $t_{\rm stop}$.

We model the massive star as an $n=2.75$ polytrope (close to $n=3$ yet stable), corresponding to $\gamma = 1+1/n = 15/11 \approx 4/3$ and an equation of state (EOS) $p = (4/11)\rho e \approx (1/3)\rho e$, where $e$ is the specific internal energy. The star has $M_\star = 100\,M_\odot$ and $R_\star \approx 16\,R_\odot$ \citep[from the main-sequence relation $R_\star = (M_\star/M_\odot)^{0.6}\,R_\odot$;][]{ToutPolsEggleton_1996MNRAS.281..257T}. 

The sBH has $M_\bullet \in [5,30]\,M_\odot$, and is placed at an initial separation $r_0 = 2R_\star$ from the stellar center. For an impact velocity at infinity $v_\infty$, the relative velocity at $r_0$ is $v = [2G(M_\star+\mbh)/r_0 + v_\infty^2]^{1/2} = (v_0^2+v_\infty^2)^{1/2}$, where $v_0^2 \equiv G(M_\star+\mbh)/R_\star$. The velocity direction is set by the impact parameter $b \equiv |\hat{\bm{v}} \times \bm{r}_0|$, with $b=0$ corresponding to a head-on collision. The specific orbital angular momentum is then $l = b\,(v_0^2+v_\infty^2)^{1/2} = b_\infty v_\infty$, where $b_\infty$ is the impact parameter at infinity. Since $b_\infty \to \infty$ for parabolic orbits ($v_\infty = 0$), we parameterize the initial conditions with $b$ rather than $b_\infty$.


\begin{figure*}
    \centering
    \includegraphics[width=\textwidth]{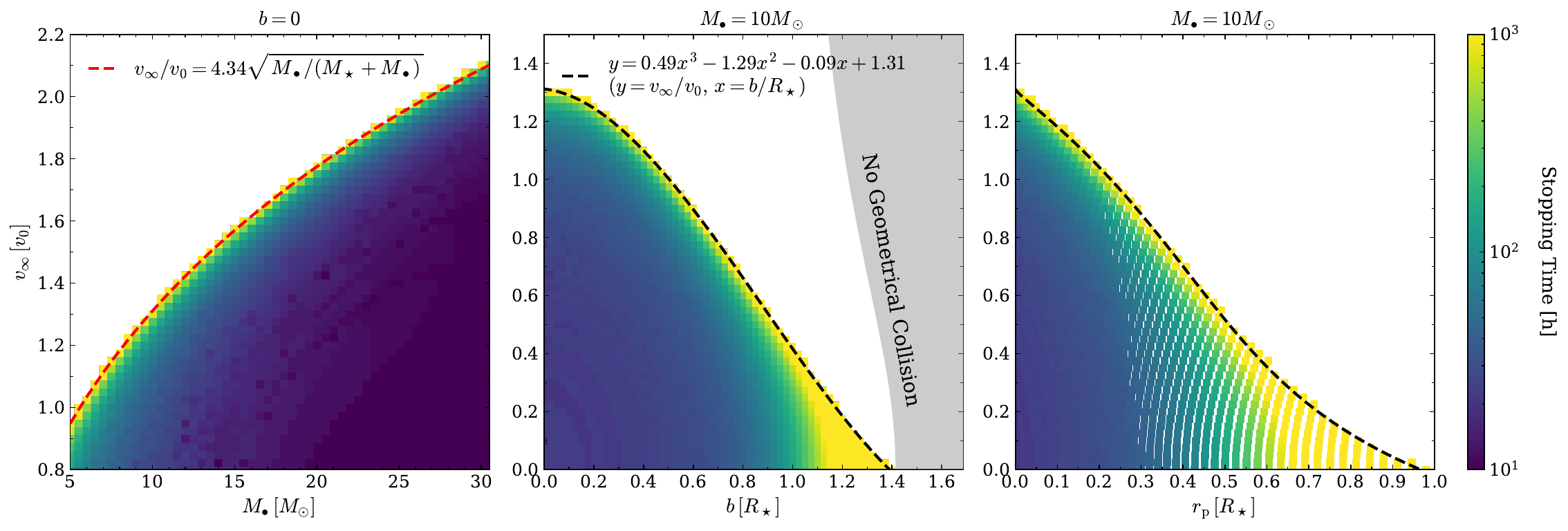}
    \caption{Critical collisional parameters for an sBH to be retained in the massive star and form a BH*, which implies a finite stopping time ($t_{\rm stop}$) through our numerical orbital integration (Section~\ref{sec:semi-analytic}).
    \emph{Left}: the stopping time in the parameter space of $(\mbh/M_\odot, v_\infty/v_0)$, assuming head-on collisions ($b = 0$). The colored mesh plot indicates the stopping time (in hours) for sBHs that successfully remain inside the star. The red dashed line marks the retention boundary. 
    \emph{Middle}: the stopping time in the parameter space of $(b/R_\star, v_\infty/v_0)$, assuming fixed $M_\bullet=10\,M_\odot$. The black dashed line approximates the retention boundary in the $b-\vinf$ plane. {The gray-shaded region indicates the parameter space where the sBH does not geometrically collide with the star.
    \emph{Right}: similar to the middle panel, but transforming the x-axis from the impact parameter $b$ to the periastron distance $r_{\rm p}$.
    }
    }
    \label{fig:critical_boundary}
\end{figure*}

\subsubsection{Critical sBH mass to form a BH*}
\label{sec:sBHmass}

Figure~\ref{fig:b-0.0} shows the orbital evolution of sBHs with $\mbh=5,\,10,\,30~\msun$ in head-on collisions ($b=0$) with $\vinf/v_0=0,\,1,\,2$. From top to bottom, the panels show the position $r$, the total velocity $v$, and the total energy $E=\mbh (v^2/2 + \Phi(r))$ of the sBH, where $\Phi(r)$ is the gravitational potential energy due to the background star,
\begin{equation}
    \Phi(r) =
\begin{cases}
 -GM_\star / r, & r \geq R_\star;\\
 -GM_\star(<r)/r - G\int_r^{R_\star} 4\pi r' \rho(r') {\rm d} r', & r < R_\star.
\end{cases}
\end{equation}
All sBHs that are eventually retained inside the star are stopped within dozens of hours, i.e., a few stellar dynamical times, consistent with the order-of-magnitude estimate in Section~\ref{sect:BHL-drag}. Most of the energy dissipation occurs near the stellar core, where the higher gas density strengthens the dynamical friction. More massive sBHs are damped faster, as expected from Equation~\eqref{equ:stoping_time}: for $\vinf/v_0=0$, the stopping time decreases from $\sim37\,\rm h$ at $\mbh=5\,\msun$ to $\sim9\,\rm h$ at $\mbh=30\,\msun$. They also deposit more energy: for $\mbh=30\,\msun$, the dissipated energy becomes comparable to the stellar binding energy ($\sim GM_\star^2/R_\star$), implying significant envelope expansion and possibly substantial mass loss.

Consequently, there exists a critical sBH mass above which the dissipative heating injects enough energy to disrupt the host star, setting a natural upper limit for an sBH to be retained and form a BH*. Our semi-analytic simulations yield a first-order scaling between this critical mass and the stellar mass. As illustrated by the gray dashed lines in the bottom panels of Figure~\ref{fig:b-0.0}, the final energy of an sBH settling at the stellar center equals its local potential energy, $\Phi(r=0) = -3.80\,GM_\star /R_\star$, with the numerical coefficient set by the polytropic index $n=2.75$. For mild velocities at infinity, the energy dissipated into the stellar envelope is $E_{\rm drag} = \mbh(\vinf^2/2 - \Phi(r=0)) \approx -\mbh\Phi(r=0)$. The binding energy of a polytropic star is $E_{\rm bind} = -\alpha(n)\, GM_\star^2/R_\star$ \citep{Chandrasekhar_1957isss.book.....C}, where $\alpha(n) = 3/[2(5-n)] \approx 0.67$ for $n=2.75$. Equating $|E_{\rm drag}|$ with $|E_{\rm bind}|$ yields a maximum sBH mass that scales linearly with the stellar mass,
\begin{equation}
    M_{\bullet,\rm max} \approx \frac{0.67 \,GM_\star^2/R_\star}{3.80\,GM_\star/R_\star} \approx 0.2\,M_\star. \label{equ:critical_mass_for_bh_star}
\end{equation}
This condition holds for massive star that can be approximated with $n\lesssim 3$. For $M_\star = 100\,\msun$, the critical sBH mass is $M_{\bullet,\rm max} \approx 20\,\msun$, so more massive sBHs (e.g., $30\,\msun$) may drive significant envelope ejection. 


\subsubsection{Critical collisional parameters to retain the BH}
\label{sec:sBHimpact}

Another important question is with what collisional parameters $(b,v_\infty)$ can the sBH be retained at the center of the star after a collision. In our example head-on ($b=0$) collisions in Figure~\ref{fig:b-0.0}, there is an emerging trend that more massive sBHs can be easier to be retained despite at a higher velocity ($v_\infty$). {To validate this trend over a larger parameter space}, we first fix the impact parameter to $b=0$, and perform a parameter survey over the sBH mass ($\mbh$) and velocity at infinity ($v_{\infty}$) to {investigate} the retention boundary. 
The results are presented in the left panel of Figure~\ref{fig:critical_boundary}, where {the blank cells} mark the $\mbh$--$v_{\infty}$ parameter space leading to an escape, while the colored mesh indicates the stopping time for retained sBHs. 
For sBHs with $\mbh \lesssim 30\,M_\odot$, the maximum velocity at infinity is constrained to $v_{\infty} \lesssim 2v_0 = 2\sqrt{G(M_\star+\mbh)/R_\star}$. Inside the retention boundary, sBHs will be damped to the star's center in $\sim 10$--$100\,\rm h$.

{The maximum velocity at infinity that permits retention, $\vinf$, generally increases with sBH mass. This trend follows from the $\mbh^2$ scaling of the gas drag force (Equation~\ref{equ:drag-force}), which allows more massive sBHs to dissipate kinetic energy more efficiently. For capture, the energy dissipated during the first passage through the star must exceed the kinetic energy at infinity, $\mbh\vinf^2/2$. The dissipated energy during one penetration is approximately
\begin{equation}
    E_{\rm drag} \sim F_{\rm HL} R_\star
\sim \frac{4\pi G^2 \rho R_\star \mbh^2}{v_0^2}.
\end{equation}
Equating these energies gives $\vinf \propto \mbh^{1/2}/v_0$ for fixed stellar properties, or $\vinf \propto \mbh^{1/2}$ when variations in $v_0$ are negligible. Motivated by this scaling and accounting for the mass dependence of $v_0$, we fit the retention boundary in the left panel of Figure~\ref{fig:critical_boundary} with
\begin{equation}
\frac{\vinf}{v_0}
= a\sqrt{\frac{\mbh}{M_\star+\mbh}}, \quad {\rm or} \quad v_\infty = a \sqrt{\frac{G M_\bullet}{R_\star}} .\label{equ:v_inf_for_capture}
\end{equation}
The best-fit value, $a \approx 4.34$, closely reproduces the boundary (red dashed line in Figure~\ref{fig:critical_boundary}). This coefficient depends on the stellar properties and density profile and must be determined numerically.
}

We note that while massive sBHs can be efficiently retained in the star, the collision may not form a BH* if the sBH mass is above the threshold ($\sim 0.2\,M_\star$, cf. Equation~\ref{equ:critical_mass_for_bh_star})

Whether the sBH can be captured also depends critically on the impact parameter $b$ {(or equivalently the periastron distance $r_{\rm p}$).} Taking a representative sBH with $\mbh=10\,M_\odot$, we conduct a grid search over $b$ and $v_{\infty}$ to map the retention threshold. 
{The results are displayed in Figure~\ref{fig:critical_boundary}, where the middle panel is in $b-\vinf$ plane and the right panel is in $r_{\rm p}-\vinf$ plane.}
{The gray-shaded region in the middle panel denotes the parameter space in which the sBH trajectory does not geometrically intersect the stellar surface, so no geometrical collision can occur.}
{For parabolic orbits ($\vinf=0$), all geometrical collisions produce sBH retention, because even a small amount of dissipation makes the sBH orbital energy negative, binding it to the star.
As $b$ decreases, the critical $\vinf$ for retention increases. This is expected, as trajectories with smaller impact parameters penetrate closer to the high-density stellar core and experience stronger dissipation (cf. Equation~\ref{equ:drag-force}).} 
Ultimately, when $v_{\infty} \sim 1.3\,v_0$, only head-on ($b=0$) collisions can marginally retain the sBH inside the star. 
{We fit a cubic polynomial to the retention boundary, and get $\vinf/v_0=0.49(b/R_\star)^3-1.29(b/R_\star)^2-0.09(b/R_\star)+1.31$ (black dashed lines in Figure~\ref{fig:critical_boundary}).}
However, for most sBHs retained in the star, the stopping time is still $\sim 10$--$100\,\rm h$.

\subsection{Dissipative heating of the star}

\begin{figure*}[t]
    \centering
    \includegraphics[width=0.9\textwidth]{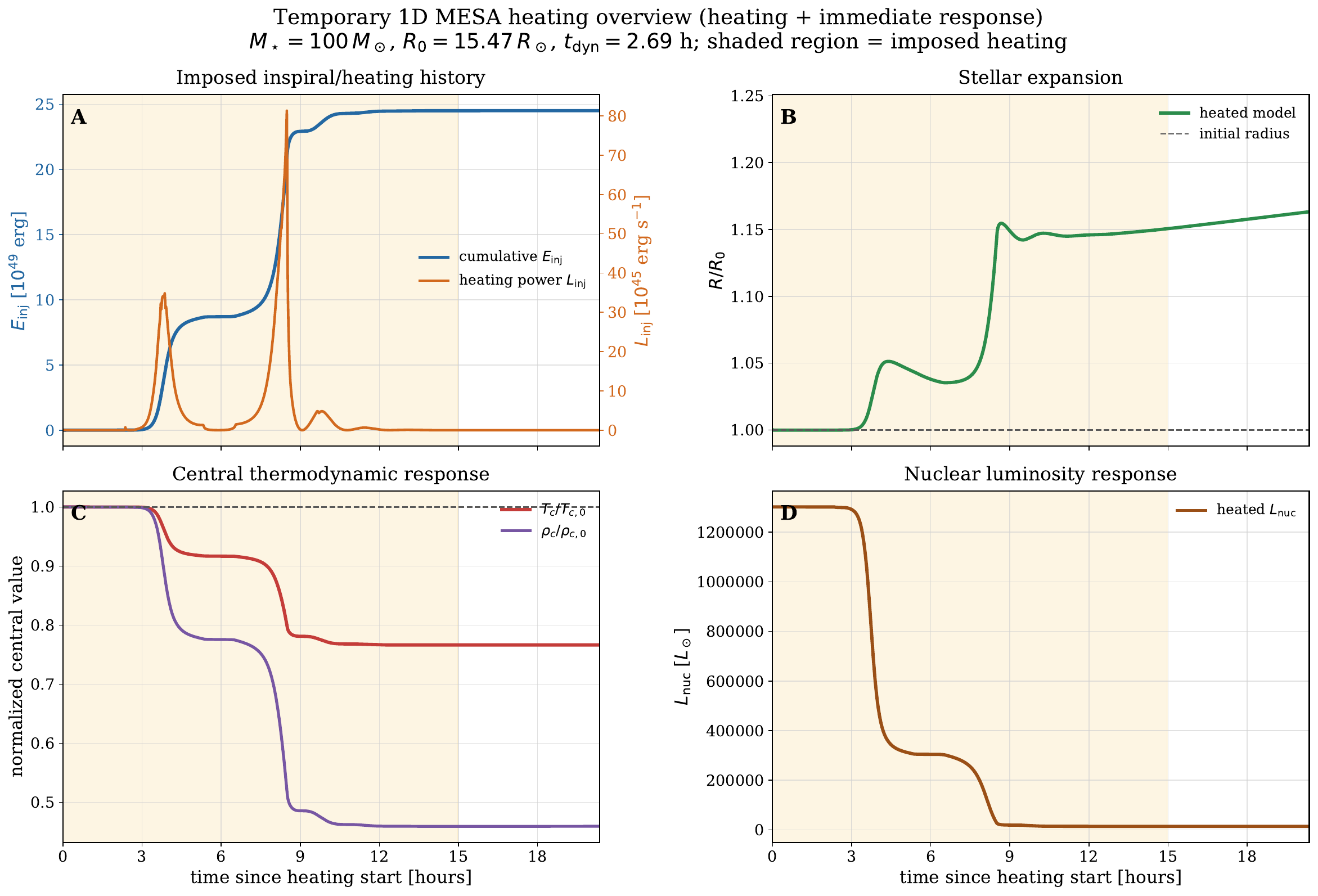}
    \caption{
    Response of the fiducial $100\,M_\odot$ stellar model to imposed inspiral heating 
    of a $10\,M_\odot$ sBH with $b=0$ and $v_\infty=0$. 
    The shaded region marks the heating interval. Panel (a) shows the cumulative injected energy and heating luminosity. 
    Panel (b)––(d) show the stellar expansion in radius, central temperature and
    density, and nuclear luminosity.
    The shaded region marks the heating interval.
    }
    \label{fig:mesa-heating}
\end{figure*}

\label{subsec:mesa-heating}
While the semi-analytic model in Section \ref{sec:sBHmass} and 
Section \ref{sec:sBHimpact} follows the orbital evolution of the sBH, it
does not include the response of the background star to the dissipated orbital energy,
which simultaneously heats the stellar interior. 
This energy injection may cause envelope expansion and mass loss, 
and additionally, it may raise the temperature of the nuclear-burning 
zone and boost the burning rate, potentially 
triggering a thermal runaway if the star cannot adjust quickly enough.
To address this concern, we use \texttt{MESA} to calculate the one-dimensional stellar
response to the dissipative heating of the star by the damping of the sBH orbit
\citep{PaxtonBildstenDotter_2011ApJS..192....3P,PaxtonCantielloArras_2013ApJS..208....4P,PaxtonMarchantSchwab_2015ApJS..220...15P,PaxtonSchwabBauer_2018ApJS..234...34P,PaxtonSmolecSchwab_2019ApJS..243...10P,JermynBauerSchwab_2023ApJS..265...15J}.
Our setup builds on previous massive AGN-star calculations with \texttt{MESA}
\citep{CantielloJermynLin_2021ApJ...910...94C,Ali-DibLin_2023MNRAS.526.5824A,
Xu_2025RAA....25k5013X,XuChenLin_2026ApJ...997..206X}.

We begin with a non-rotating main–sequence stellar model with
$M_\star=\,100 M_\odot$ and $R_\star \approx 16 R_\odot$.
The orbital energy dissipated by the sBH is  added through
the \texttt{other\_energy} hook as an external
specific heating term.
The total additional energy is given as
\begin{equation}
    E_{\rm inj}=f\frac{G M_\star^2}{R_\star}.
    \label{eq:mesa-heating-energy}
\end{equation}
The time dependence of the additional heating is taken from the calculation in
Section~\ref{subsect:orb-integ} as well as in Figure~\ref{fig:b-0.0}, 
which records the sBH's position $r(t)$ and the corresponding dissipated ${E}_{\rm inj}(t)$. 
We assume the dynamical-friction luminosity is proportional 
to the energy dissipation and linearly interpolate the curve. 
We damp the energy to stellar mesh cells between $q_{\min}$ and $q_{\max}$
where $q$ is the normalized radial distance. 
At each timestep, the interpolated luminosity is converted to a specific heating rate 
and is applied to the injected stellar cells between $q_{\min}$ and $q_{\max}$. 
With this additional source term, \texttt{MESA}
continues its calculations of stellar structure 
and evolution.

The evolution followed here is much shorter than the characteristic growth time of
the Eddington–limited sBH, so its mass changes negligibly. Moreover, the dynamical friction luminosity is much larger than the Eddington-limited accretion luminosity of the sBH, 
making accretion feedback energetically subdominant during the inspiral.
We evolve the model to $50\,t_{\rm dyn}$ after the onset of heating.

Figure~\ref{fig:mesa-heating} shows the fiducial stellar response. 
For the fiducial calculation, we use the luminosity history of a
$10\,M_\odot$ sBH with $b=0$ and $v_\infty=0$. The total injected energy is $0.1\,GM_\star^2/R_\star=6.79\times10^{49}\,{\rm erg}$.
The heating lasts
$5.39\times10^4\,{\rm s}=14.97\,{\rm h}=5.6\,t_{\rm dyn}$.

The radius increases from $15.47\,R_\odot$ to $17.95\,R_\odot$ during
the heating phase and reaches $19.18\,R_\odot$ by the end of the calculation.
The expansion is accompanied by decreases of approximately $15\%$ in
$T_c$ and $35\%$ in $\rho_c$. The nuclear luminosity falls from
$1.30\times10^6\,L_\odot$ to $1.37\times10^4\,L_\odot$ by the end of
heating. The model therefore expands and cools without developing a nuclear
runaway over the computed interval.

The star's expansion is accompanied by decreases of approximately $15\%$ in $T_{\rm c}$ and $35\%$ in $\rho_{\rm c}$, consistent with the scalings $T_{\rm c}\propto R_\star^{-1}$ and $\rho_{\rm c}\propto R_\star^{-3}$.  
The nuclear luminosity decreases more strongly because of the steep temperature dependence of CNO burning near $T_{\rm c}\simeq2\times10^7\,{\rm K}$ 
\citep{PaxtonCantielloArras_2013ApJS..208....4P,PaxtonMarchantSchwab_2015ApJS..220...15P}. 
Thus, the model expands and cools rather than developing a nuclear runaway over the computed interval.


We also vary the injected-energy fraction, expressed relative to the stellar
gravitational-energy scale in Equation~\ref{eq:mesa-heating-energy},
and the damping time while keeping
$M_\star=100\,M_\odot$ and $0\leq q\leq0.8$ fixed. 
The completed models cover
$f = 0.03$, $0.1$, $0.3$ or $0.5\,GM_\star^2/R_\star$.
The calculations with $f=0.5$ terminate before the end of
the heating phase because of excessive expansion.

These one-dimensional models isolate the thermal response to orbital energy
dissipation. The wake, shocks, angular-momentum deposition, and dynamical
mass loss are followed in the hydrodynamical simulations in
Section~\ref{sec:hydro}.


\section{Hydrodynamical Simulations}
\label{sec:hydro}
The results in Section \ref{subsec:mesa-heating} indicate that energy dissipation
during the sBH's orbital decay can lead to changes in the massive main-sequence star's interior.  However, this 1-D \texttt{MESA} model does not fully capture the rapid
3-D dynamical response of the stellar envelope.  In this section, 
we perform a suite of hydrodynamical simulations of star--sBH collisions, using the Lagrangian code \texttt{GIZMO} \citep{Hopkins_2015MNRAS.450...53H}. The simulations solve the standard hydrodynamical equations\footnote{Specifically: (1)~$\partial_t \rho + \nabla \cdot ( \rho \bm{v})  = 0$; (2)~$\partial_t (\rho \bm{v})  + \nabla \cdot ( \rho \bm{v} \bm{v} + p \mathbf{I})  = \rho \bm{g}$; (3)~$\partial_t \left( \rho e+ \rho v^2/2\right)  + \nabla \cdot \left[\left(\rho e+\rho v^2/2 + p \right) \bm{v} \right] = \rho \bm{g}\cdot \bm{v}$; (4)~$p = (\gamma-1)\rho e$.} with the code's Meshless Finite-Mass (MFM) mode, and self-gravity is solved with tree-gravity \citep[][]{HopkinsNadlerGrudic_2023MNRAS.525.5951H}. 

Each initial condition for the simulations includes a massive star and a companion BH (cf. Section~\ref{sect:initial-condidtions}). The star is fixed as an $n=2.75$, $100\,M_\odot$ polytrope with $128^3$ equal-mass gas cells ($\Delta m \sim 5\times 10^{-5}\,M_\odot$), which is generated following \citet{OhlmannRopkePakmor_2017A&A...599A...5O} and \citet{ShiHuangLin_2026arXiv260800242S}. {While the stellar profile modeling may influence the simulation results, we find good overall consistency between different $n$'s, which is presented in Appendix~\ref{app:tests}. In the same appendix, we also prove the resolution convergence of the hydrodynamical simulations. 
}



\begin{deluxetable}{rllll}
\tablewidth{0pt} 
\tablecaption{{Parameter space of the hydrodynamical simulations. We set four different sBH 
masses ($M_\bullet$). For each BH mass, we set a variety of orbits that are determined by the impact parameter ($b$, defined in Section \ref{sect:initial-condidtions}, which is \emph{not} the impact parameter at infinity) and the relative velocity at infinity ($v_\infty$). The periapsis ($r_{\rm p}$) of each orbit is also derived.
Here $v_0^2 \equiv G(M_\star+M_\bullet)/R_\star$. The 
non-spinning stars are fixed with $M_\star =100\,M_\odot$ and $R_\star=100^{0.6}\,R_\star$ for all simulations, and are modeled with $n=2.75$ polytropes by default.}
}
\label{tab:ics}
\tablehead{\colhead{$M_\bullet\,[M_\odot]$} & $v_{0}\,[\rm km\,s^{-1}]$ & \colhead{$b\,[R_\star]$} & \colhead{$v_{\infty}\,[v_{0}]$} & $r_{\rm p}\,[R_\star]$ }
\startdata
5 & 1124 & 0 & 0, 1, 2 &  0, 0, 0  \\
5 & 1124 & 0.5 & 0, 1, 2 &  0.125, 0.225, 0.362  \\
10 & 1151 & 0 & 0, 1, 2 &  0, 0, 0  \\
10 & 1151 & 0.5 & 0, 1, 2 &  0.125, 0.225, 0.362  \\
30 & 1251 & 0 & 0, 1, 2 &  0, 0, 0  \\
30 & 1251 & 0.5 & 0, 1, 2 &  0.125, 0.225, 0.362  \\
100 & 1552 & 0 & 0, 1, 2 &  0, 0, 0  \\
100 & 1552 & 0.5 & 0, 1, 2  &  0.125, 0.225, 0.362
\enddata
\end{deluxetable}

\subsection{Simulation setups}

\begin{figure*}
    \centering
    \gridline{
        \fig{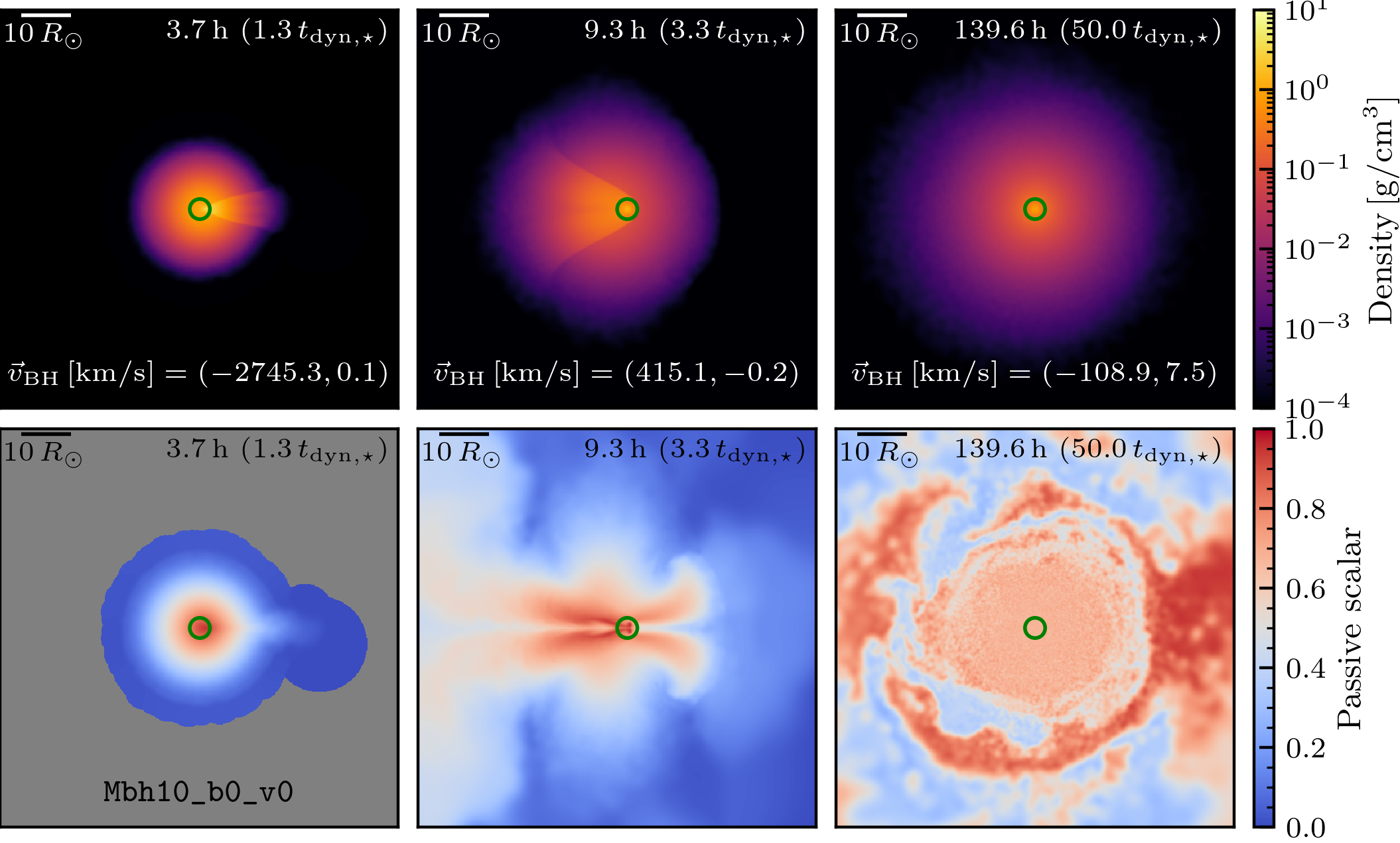}{0.5\textwidth}{(a) BH mass $10\,M_\odot$. Head-on ($b=0$) collision with $v_\infty=0$. Panels in the right column are centered at the sBH's location. }
        \fig{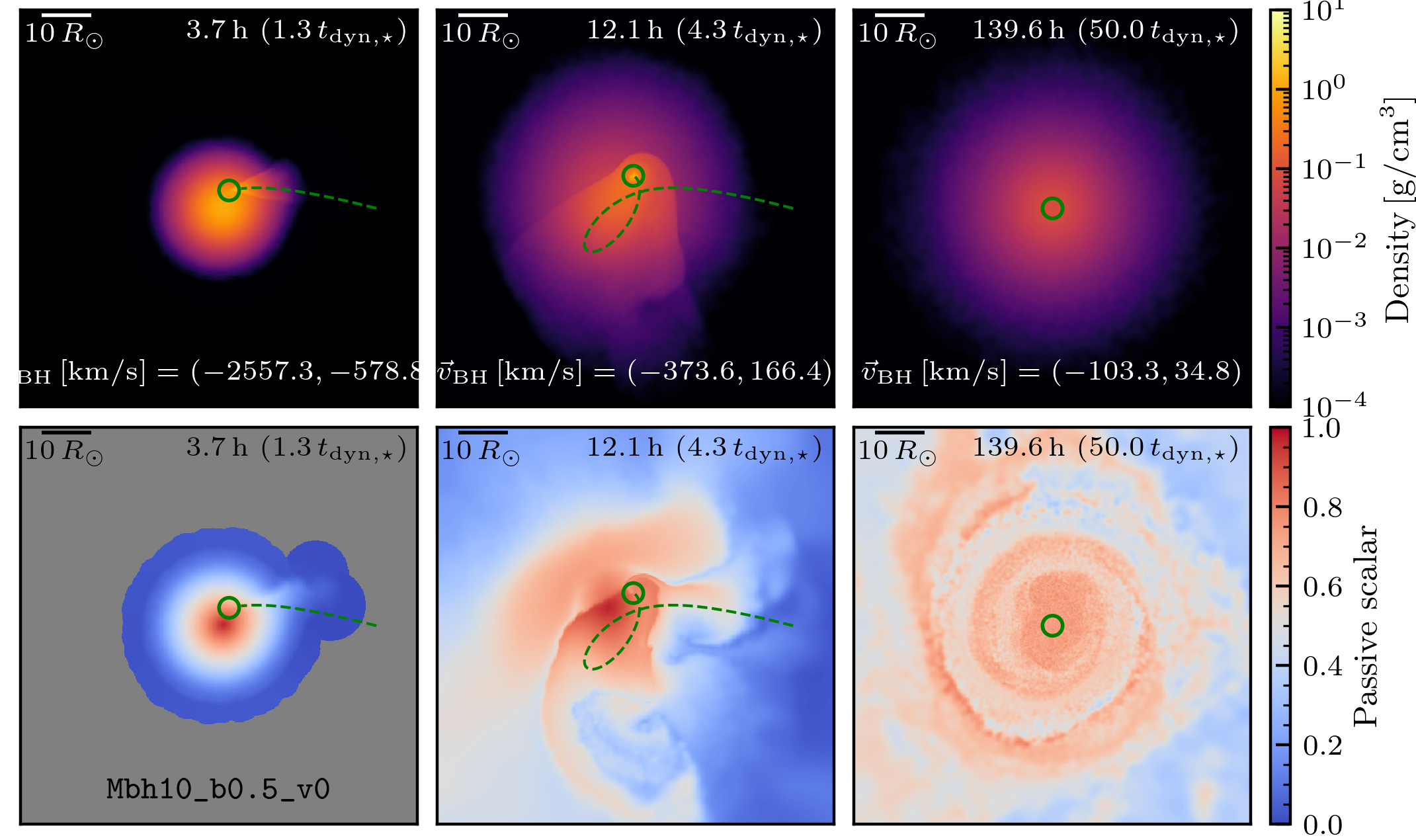}{0.5\textwidth}{(b) BH mass $10\,M_\odot$. {Off-center ($b=0.5\,R_\star$, $r_{\rm p}=0.125\,R_\odot$)} collision with $v_\infty=0$. Panels in the right column are centered at the sBH's location. }
    }
    \gridline{
        \fig{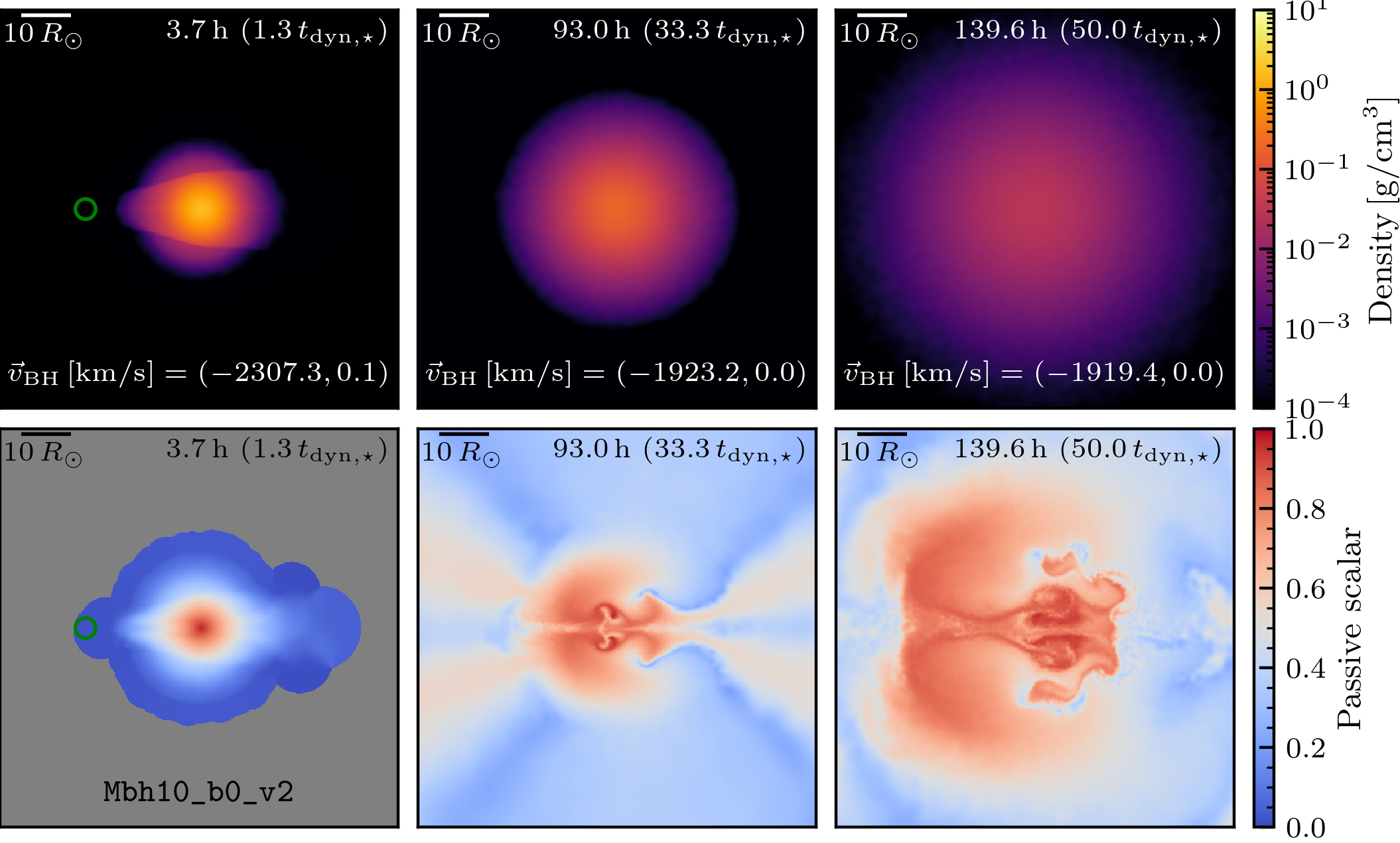}{0.5\textwidth}{(c) BH mass $10\,M_\odot$. Head-on ($b=0$) collision with $v_\infty=2 \sqrt{G(M_\star+M_\bullet)/R_\star}$. Panels in the right two columns are centered at the star. }
        \fig{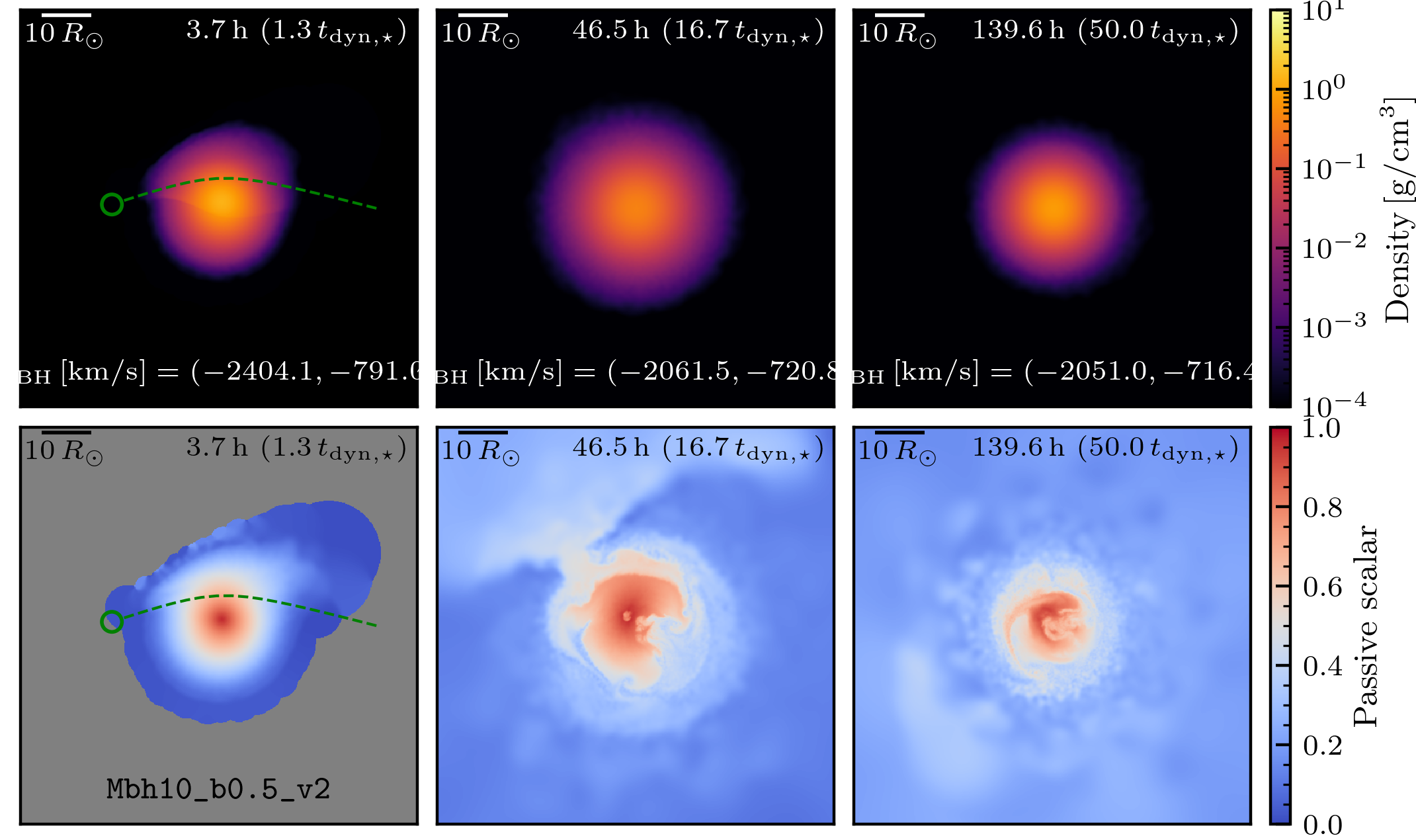}{0.5\textwidth}{(d) BH mass $10\,M_\odot$. {Off-center ($b=0.5\,R_\star$, $r_{\rm p}\approx 0.362\,R_\odot$)} collision with $v_\infty=2 \sqrt{G(M_\star+M_\bullet)/R_\star}$. Panels in the right two columns are centered at the star.}
    }
    \gridline{
        \fig{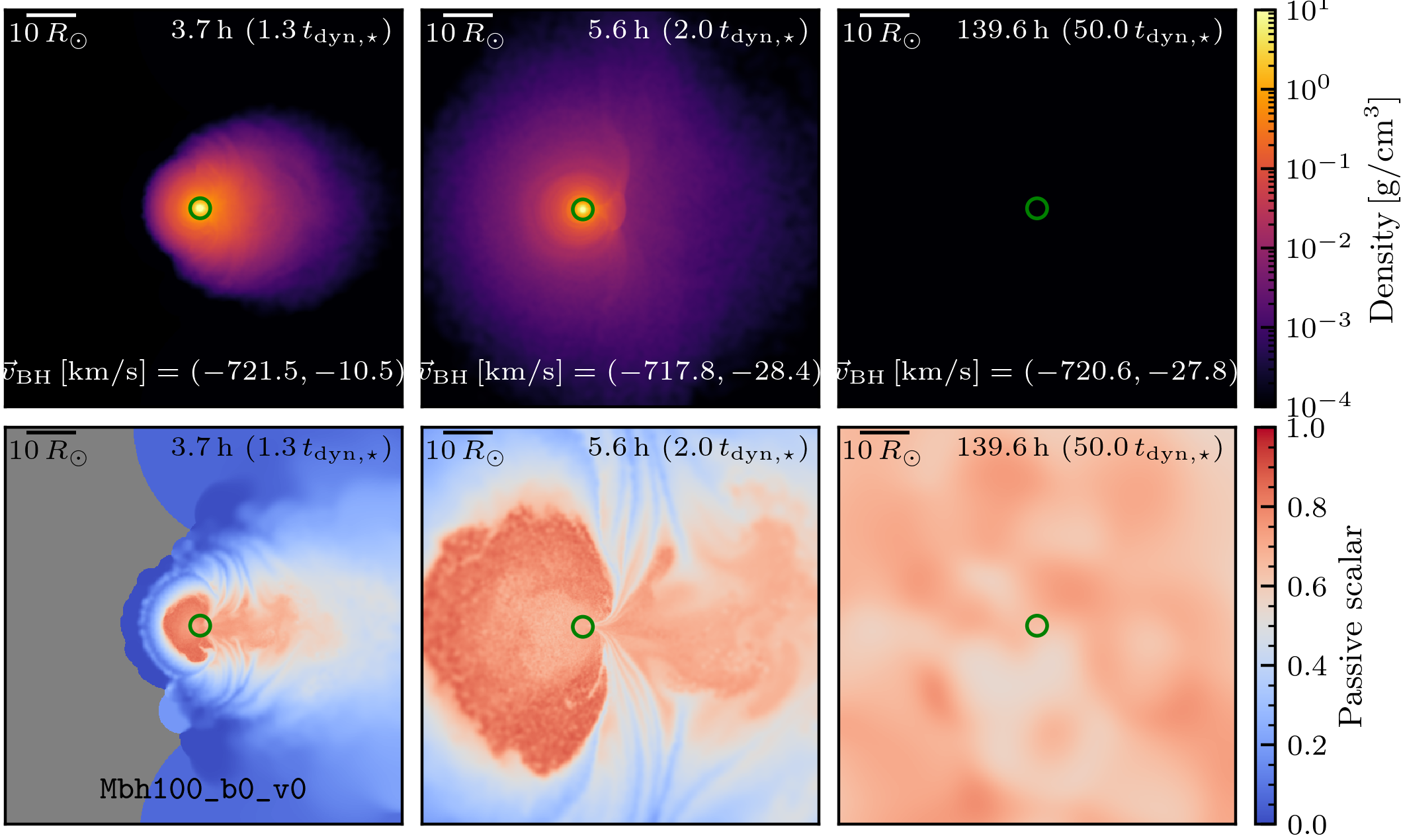}{0.5\textwidth}{(e) BH mass $100\,M_\odot$. Head-on ($b=0$) collision with $v_\infty=0$. Panels in the right column are centered at the sBH's location.  }
        \fig{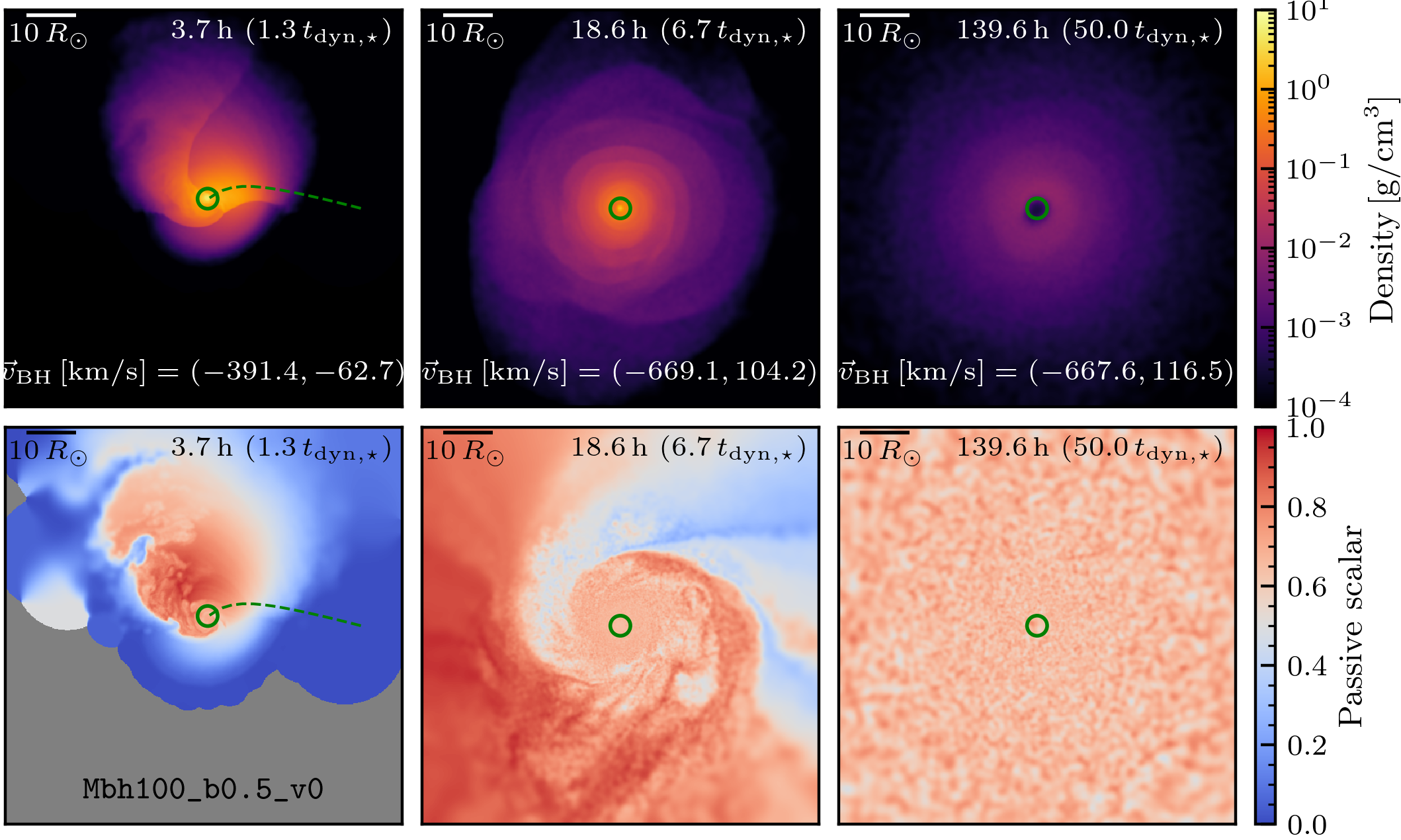}{0.5\textwidth}{(f) BH mass $100\,M_\odot$. {Off-center ($b=0.5\,R_\star$, $r_{\rm p}=0.125\,R_\odot$)} collision with $v_\infty=0$. Panels in the right two columns are centered at the sBH's location. }
    }
    \caption{Visualization of a subset of our simulations. In each sub-figure, we show the distribution of density (\emph{top panels}) and passive scalar (\emph{bottom panels}) at the mid-plane ($z=0$), and the sBH's location (\emph{green circle}), for snapshots at different evolutionary stages (as labeled in each panel). We default to the lab frame unless otherwise mentioned. We also label the velocity of the sBH in the lab frame. For {off-center} collisions ($b\neq 0$), we plot the trajectory of the sBH (\emph{green dashed}).  }
    \label{fig:vis_fiducial}
\end{figure*}

The simulation setups are summarized in Table~\ref{tab:ics}. In each pair of $(M_\star, M_\bullet)$, we set different collisional parameters (as detailed in Section~\ref{sect:initial-condidtions}): $b/R_\star = 0, 0.5$, and $v_\infty/v_0 = 0, 1, 2$. 
This covers cases like head-on vs. {off-center} collisions, and parabolic vs. hyperbolic orbits. {Table~\ref{tab:ics} also shows the periapsis ($r_{\rm p} = l^2/(GM + \sqrt{G^2M^2+l^2 v_\infty^2})$, where $M=M_\star+M_\bullet$) for each orbit configuration, which directly measures how close the orbit can be to the star's center.} Each simulation runs for $50\,\tau_{\rm dyn,\star} \approx 140\,\rm h$ (where $\tau_{\rm dyn,\star}=[R_\star^3/(GM_\star)]^{1/2}$).
We label each simulation with \verb|Mbh%g_b%g_v%g|, representing the BH mass ($M_\bullet/M_\odot$), impact parameter ($b/R_\star$), and initial relative velocity at infinity ($v_{\infty}/v_0$). 

For gravity, we use fixed, small ($h_{\rm gas}=10^{-4}\,R_\odot$) softening radii for the gas cells, and a larger softening radius ($h_{\rm BH}=0.5\,R_\odot$) for the sBH, which is compatible with the spatial resolution. This avoids the noisy, unphysical numerical heating of the gas when the BH softening radius is too small, an issue that appeared in many of our background tests. Moreover, the Bondi radius of a $5\,M_\odot$ sBH embedded in a $c_{\rm s}\approx 1000\,\rm km\,s^{-1}$ is $R_{\rm Bondi}=GM_\bullet/c_{\rm s}^2\sim R_\odot$; given our mean spatial resolution of $\Delta r \sim 2R_\star/128 \sim 0.2\,R_\odot$, $h_{\rm gas} \ll \Delta r < h_{\rm BH} \lesssim R_{\rm Bondi}$, suggesting that the Bondi radius is resolved. {One caveat is, from our simulations, the BH--gas relative velocity can reach $\sim3000\,\rm km\,s^{-1}$ (see Figure~\ref{fig:b-0.0}), 
the HL radius ($R_{\rm HL} = 2 GM_\bullet/v^2$) is $\sim 0.2\,R_\odot$ for a $5\,M_\odot$ sBH, which is smaller than $h_{\rm BH}$. However, the HL radius is better resolved for more massive ($\geq 10\,M_\odot$) BHs.  }

Our simulations solve only the coupled hydrodynamics and gravity of the star--BH collision. We neglect BH accretion, feedback, and radiation transfer, since our simulation timescale is short ($\sim 140\,\rm h$): for Eddington-limited BH accretion, the relevant timescale is the Salpeter time $\tau_{\rm Sal} = 0.1 \sigma_{\rm es} c /(4\pi G m_{\rm p}) \approx 45\, \rm Myr$; the thermal (Kelvin-Helmholtz) timescale for a $100\,M_\odot$ star is $\tau_{\rm KH} \sim GM_\star^2/(2R_\star L_\star)\sim 10^4\,\rm yr$ (assuming near-Eddington luminosity). Both the accretion and thermal timescales are substantially longer than our simulation timescale (i.e., $\tau_{\rm dyn,\star} \ll \tau_{\rm KH} \ll \tau_{\rm Sal}$). Additionally, \citet{Murguia-BerthierMacLeodRamirez-Ruiz_2017ApJ...845..173M} found that accretion disks can hardly form in common envelopes, especially in high-density, low-compressibility regions; strong feedback is therefore uncommon, suggesting feedback can also be unimportant.


\begin{figure*}
    \centering
    \includegraphics[width=\linewidth]{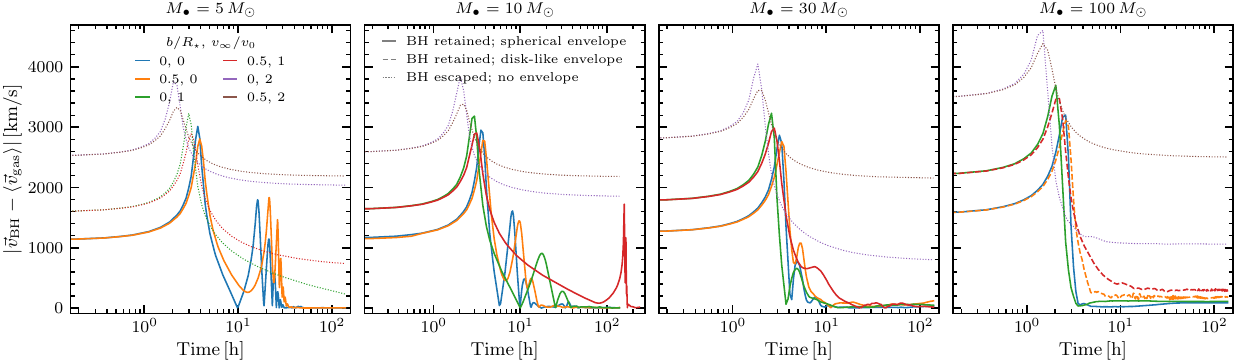}
    \caption{The relative velocity between the BH and star. This quantity is evaluated as $|\bm{v}_{\bullet}-\langle \bm{v}_{\rm gas} \rangle|$, where $\langle \bm{v}_{\rm gas} \rangle$ is the center-of-mass velocity for gas. Each panel represents the collisional simulations of different BH masses (as labeled above), where we plot the BH-gas relative velocity for collisions with varying $b$ and $v_\infty$. The possible outcome of each collision is classified as either a BH* with a spherical envelope (\emph{solid}), a BH with a disk-like envelope (\emph{dashed}), or a ``bare'' BH without an envelope (\emph{dotted}).
    }
    \label{fig:bh_vel_evo}
\end{figure*}

\subsection{Outcomes of the collisions}

\subsubsection{Morphology}

Figure~\ref{fig:vis_fiducial} visualizes a subset of the star--sBH collisional simulations, where $M_\bullet = 10\,M_\odot$, with varying impact parameters and velocities. Each sub-figure displays the density (\emph{upper panel}) and passive scalar (\emph{lower panel}) distributions at the stellar mid-plane. The passive scalar is a post-processed quantity assigned to each gas cell to track mixing during the collision; specifically, its value at $t=0$ is defined as $p=(R_\star-r)/R_\star$, where $r$ is the distance from the stellar center and $R_\star$ is the stellar radius. While the $p$-value of each gas cell is advected without change, the overall passive scalar distribution evolves as a result of mixing.

Figure~\ref{fig:vis_fiducial}a shows the head-on collision ($b=0$, $v_\infty=0$) at characteristic times $t=3.7\,\rm h$, the BH enters the star with a relative velocity of $2745\,\rm km\,s^{-1}$, driving a shock wake. Gas drag then decelerates the BH and reverses its direction of motion; by $t=9.3\,\rm h$, the BH is moving in the direction opposite to its initial trajectory. Finally, at the end of the simulation ($t=140\,\rm h$), the BH is fully decelerated and co-moves with the stellar envelope. The passive scalar distribution traces the accompanying perturbation of the stellar interior.

Figure~\ref{fig:vis_fiducial}b shows an {off-center} collision ($b=0.5\,R_\star$, $v_\infty=0$). The non-zero angular momentum causes the BH to orbit the center of the gravitational potential (see green dashed lines for its trajectory). 
As in the head-on case, the BH velocity is reversed by $t=12\,\rm h$, and the BH is fully damped by the end of the simulation.

When the initial velocity is sufficiently high, however, the BH can penetrate and escape the star, with a post-encounter speed exceeding the local escape velocity. Figure~\ref{fig:vis_fiducial}c and \ref{fig:vis_fiducial}d illustrate this regime: for both head-on and {off-center} geometries, the BH escapes when $v_\infty = 2v_0 \approx 2302\,\rm km\,s^{-1}$. The stellar interior is strongly mixed along the shock wake, yet the star remains gravitationally bound and undergoes sustained radial oscillations in response to the passing BH's perturbation.

Figures~\ref{fig:vis_fiducial}e and \ref{fig:vis_fiducial}f present the same suite of collisions for a $100\,M_\odot$ BH, equal to the stellar mass. Compared with the $10\,M_\odot$ case, the more massive BH experiences a stronger drag force (cf.\ Equation~\ref{equ:drag-force}, where $F_{\rm drag}\propto M_\bullet^2$) and exerts a deeper gravitational potential well, attracting more ambient gas. 
The BH is not represented by a sink-cell, i.e. we do not
take into account of its accretion and the associated radiative feedback.  
As an approximation, we modify BH's gravitational potential with an 
arbitrarily-specified softening length equals $0.5 R_\odot$ such that
BH's gravity would be suppressed in its close proximity.
In the head-on cases, the BH accumulates a gas envelope even during the high-velocity collision at $v_\infty=2v_0\approx3104\,\rm km\,s^{-1}$. However, 
this envelope is dynamically disrupted following the violent shock interaction, and in both cases it is stripped by the end of the simulation.

The {off-center} geometry produces qualitatively different behavior. At $v_\infty=0$ and $b=0.5\,R_\star$ (Figure~\ref{fig:vis_fiducial}f), the angular momentum of the encounter causes the star to deform and spiral around the $100\,M_\odot$ BH, ultimately forming a rotationally supported, disk-like 
envelope 
by $t=19\,\rm h$. 
{The buildup of high temperature and pressure gradient 
in the disk around the BH's softened potential marginally suppresses 
gravitational instability and the excitation of global spiral structure.}
Although the disk loses mass continuously and is susceptible to 
marginal dynamical instabilities, it survives until the end of the simulation. At high collisional velocity ($v_\infty=2v_0\approx3104\,\rm km\,s^{-1}$; not shown here), the BH escapes before a disk can form, and the star is heavily disrupted.

\subsubsection{BH retention in the star}\label{sec:bh_retention}

Figure~\ref{fig:bh_vel_evo} quantifies whether the BH is efficiently damped by gas drag and retained within the star, using the relative velocity between the BH and the stellar center of mass throughout the evolution, $|\bm{v}_\bullet - \langle \bm{v}_{\rm gas} \rangle|$, where $\langle \bm{v}_{\rm gas} \rangle$ is the mass-weighted center-of-mass velocity of all gas cells (note that our simulations employ equal-mass cells). The relative velocity at $t=0$ corresponds to the initial separation of $2\,R_\star$ and is distinct from $v_\infty$.

For most simulations, the relative velocity is damped to $\sim 0\,\rm km\,s^{-1}$ within $\sim 100\,\rm h$; for the remainder it is not. This is broadly consistent with semi-analytic simulations (Figures~\ref{fig:b-0.0}, \ref{fig:critical_boundary}). Yet hydrodynamical simulations find more outcomes in terms of the gas morphology, which we classify into {four} cases.

\begin{enumerate}
    \item \emph{The BH is captured in a spherical, dense, BH*-like envelope.} The BH velocity is efficiently damped, and the BH settles at the stellar center surrounded by a dense ($\sim 1\,\rm g\,cm^{-3}$) envelope (cf.\ Figures~\ref{fig:vis_fiducial}a and ~\ref{fig:vis_fiducial}b). As shown in Figure~\ref{fig:bh_vel_evo}, this outcome is typical for low-velocity ($v_\infty/v_0\lesssim 1$) collisions with low-mass ($M_\bullet \lesssim 10\,M_\odot$) BHs. More massive BHs can also acquire an envelope, but only in head-on, low-velocity ($v_\infty/v_0\lesssim 1$) collisions.

    \item \emph{The BH is captured in a micro-TDE-like, non-rotating envelope.} This happens for head-on collisions with a massive BH: although the gas drag damps the BH, the surrounding gas envelope is disrupted. The envelope is also non-rotating since the system has no net angular momentum.

    \item \emph{The BH is captured in a micro-TDE-like, rotating envelope.} The BH is massive enough to heavily deform the star, and the orbital angular momentum drives the formation of a disk-like envelope (cf.\ Figure~\ref{fig:vis_fiducial}f). 
    This outcome occurs for low-velocity ($v_\infty/v_0\lesssim 1$), {off-center} collisions with high-mass ($M_\bullet \gtrsim 100\,M_\odot$) BHs. 

    \item \emph{The BH escapes without forming an envelope.} The BH impacts the star at sufficiently high velocity that gas drag is unable to halt it (cf.\ Figures~\ref{fig:vis_fiducial}c and \ref{fig:vis_fiducial}d). This outcome occurs for high-velocity ($v_\infty/v_0\gtrsim 2$) collisions. An exception arises for light BHs (e.g., $5\,M_\odot$), whose weaker drag force (cf.\ Equation~\ref{equ:drag-force}) allows escape even at moderate velocity ($v_\infty/v_0=1$).
\end{enumerate}

\begin{figure*}
    \centering
    \includegraphics[width=\linewidth]{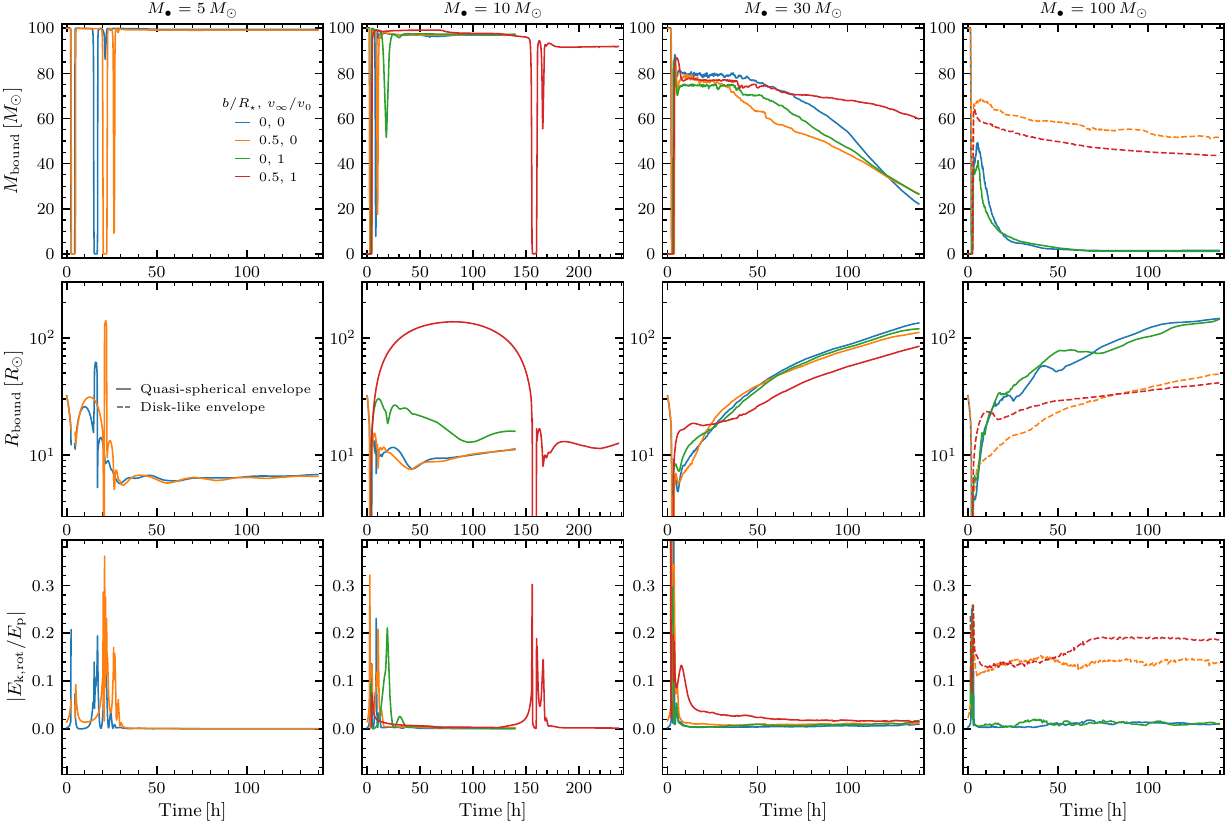}
    \caption{Envelope mass (\emph{upper panels}), half-mass radius (\emph{middle panels}), and $|E_{\rm k,rot}/E_{\rm p}|$ (\emph{bottom panels}) of the BH* system for different collisional experiments. Here $|E_{\rm k,rot}/E_{\rm p}|$ represents the ratio between the rotational kinetic energy and the potential energy for the gaseous envelope. Each column corresponds to the labeled BH mass. Solid and dashed lines distinguish quasi-spherical and disk-like envelopes, respectively.}
    \label{fig:bound_mass_radius}
\end{figure*}

For cases in which the BH is retained within the stellar material, we examine the envelope mass and radius in Figure~\ref{fig:bound_mass_radius}. The envelope mass is defined as the gas gravitationally bound to the BH$+$envelope system, i.e., satisfying $\Phi + \gamma e + |\bm{v}-\bm{v}_\bullet|^2/2 < 0$, where $\Phi$ is the gravitational potential and $\gamma e$ is the specific enthalpy; the half-mass radius is evaluated accordingly. A quasi-hydrostatic BH* is only achieved for $M_\bullet \lesssim 10\,M_\odot$, in which case the envelope retains most of the original stellar mass and the half-mass radius remains approximately steady (modulo radial oscillations). For more massive BHs ($M_\bullet \gtrsim 30\,M_\odot$), the envelope loses mass while expanding, reaching at most $\lesssim 50\,M_\odot$, and does not attain a quasi-hydrostatic state within the simulation time.

In the bottom panels of Fig.~\ref{fig:bound_mass_radius}, we show the time evolution of the ratio $|E_{\rm k,rot}/E_{\rm p}|$ (ratio between the rotational kinetic energy and the potential energy), as an indicator of whether the envelope is rotationally dominated. Both quantities are derived from gas particles in the simulations, where $E_{\rm k,rot} = \sum_i m_i v_{{\rm rot}, i}^2/2$, where $v_{{\rm rot},i}$ is the rotational (tangential) velocity in the $xy$ plane for each gas particle, which is defined as $v_{{\rm rot},i} = j_{z, i}/r_{\rm 2D} = (x v_y - y v_x)_i/\sqrt{x_i^2+y_i^2}$. The potential energy is $E_{\rm p} = \sum_i m_i \Phi_i$. For most collisions, $|E_{\rm k,rot}/E_{\rm p}|\sim 0$ at the end of the simulations; however, for $M_\bullet = 100\,M_\odot$ and $b>0$, $|E_{\rm k,rot}/E_{\rm p}|\sim 0.1$--0.2. This quantifies that a rotation-supported, disk-like envelope forms for massive sBHs with $M_\bullet = 100\,M_\odot$.

\subsubsection{BH* radial structure}

\begin{figure}
    \centering
    \includegraphics[width=\linewidth]{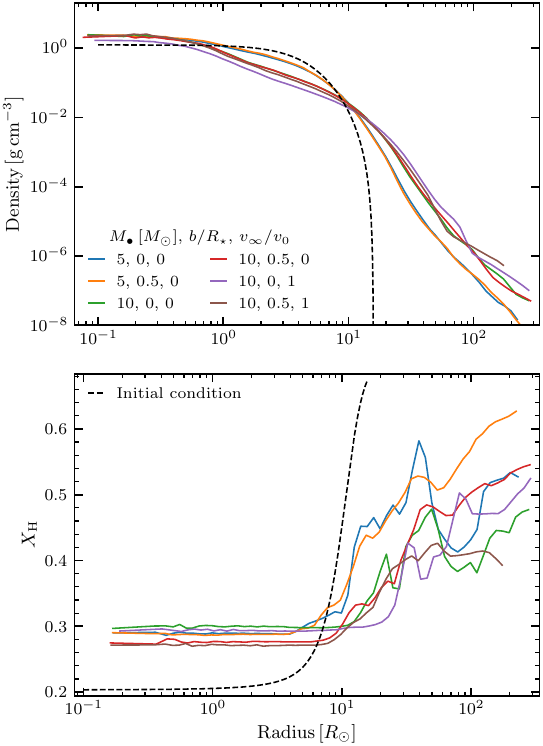}
    \caption{Radial density (\emph{upper panel}) and mock hydrogen abundance ($X_{\rm H}$; \emph{lower panel}) profiles of the BH*s produced in this work. Black dashed lines show the initial conditions. The $X_{\rm H}$ profile is a mock pre-collision distribution, post-processed to obtain the final abundance after mixing.}
    \label{fig:density_mixing}
\end{figure}

Of all collisional simulations, only 6 produce quasi-hydrostatic BH*s; these are examined further here. Figure~\ref{fig:density_mixing} shows their radial density profiles. Relative to the initial main-sequence star (modeled as a $100\,M_\odot$, $n=2.75$ polytrope), the gas envelope of the BH* is substantially more extended following the collision, while the central gas density in the vicinity of the BH remains $\sim 1\,\rm g\,cm^{-3}$, comparable to the pre-collision value.

We also quantify the redistribution of chemical elements by adopting a mock pre-collision hydrogen abundance profile motivated by nuclear burning: $X_{\rm H} = 0.2 + 0.5/[1+\exp\!\left(-(r-r_{\rm c})/2\right)]$, where $r_{\rm c} = 8\,R_\odot$. Post-processing the simulations by advecting the cells yields the final radial $X_{\rm H}$ distribution. Compared with the initial profile, the central hydrogen abundance is elevated to $\sim 0.3$ from 0.2. However, whether this hydrogen replenishment will extend the nuclear-burning time remains uncertain, since the BH* may be {partially} powered by the central accreting BH 
rather than {entirely by} nuclear reactions \citep[][]{BegelmanRossiArmitage_2008MNRAS.387.1649B}.

\section{Discussion}
\label{sec:implications}

{

\subsection{Comparing semi-analytic and hydrodynamical simulations}

\begin{figure*}
    \centering
    \includegraphics[width=\linewidth]{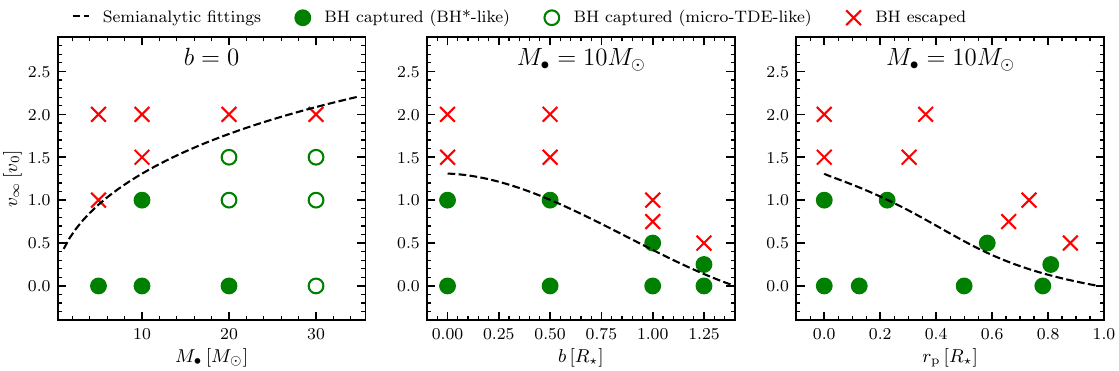}
    \caption{
    Comparison between the semi-analytic predictions and hydrodynamical simulations for the outcomes of star--sBH collisions. The dashed lines show the fitting relations obtained from the semi-analytic orbital integrations in Figure~\ref{fig:critical_boundary}, which delineate the critical boundary for retaining the sBH within the massive star. The symbols indicate the outcomes of the hydrodynamical simulations: BH*--like objects are shown by filled circles, micro-TDEs by open circles, and escaped BHs by crosses. 
    }
    \label{fig:region_plot_capture}
\end{figure*}

Figure~\ref{fig:region_plot_capture} compares the outcomes of the hydrodynamical simulations with the predictions from our semi-analytic orbital integrations. In Figure~\ref{fig:critical_boundary}, we use the semi-analytic model to determine the critical collisional parameters separating encounters in which the sBH is retained by the massive star from those in which it ultimately escapes. We further derive fitting relations for these critical boundaries as a function of the sBH mass and the initial orbital parameters. Figure~\ref{fig:region_plot_capture} provides a direct test of these relations against the full hydrodynamical calculations, in which the stellar response and the back-reaction of the gas on the sBH are treated self-consistently.

To populate the parameter space more densely, we perform an additional suite of lower-resolution ($64^3$, see the convergence tests in Appendix~\ref{app:tests}) hydrodynamical simulations spanning a range of $(M_\bullet,b,v_\infty)$. We find good overall agreement between the semi-analytic predictions and the hydrodynamical results: the fitting relations approximately separate the simulations according to whether the sBH remains bound to the star or escapes after the encounter.

\subsection{Formation of the disk-like envelope}

\begin{figure}
    \centering
    \includegraphics[width=\linewidth]{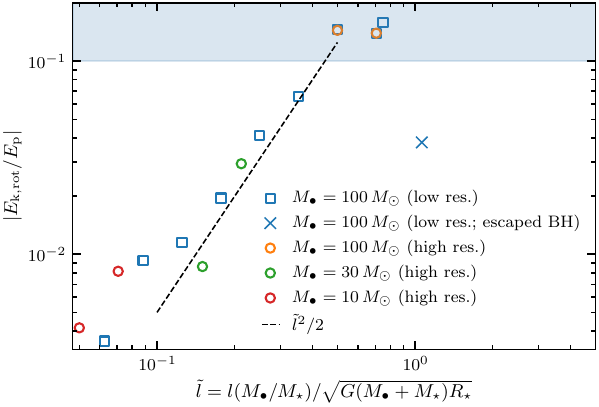}
    \caption{ Relation between the dimensionless BH angular momentum $\tilde{l}$ and the rotational energy of the remnant stellar envelope. Squares and circles show simulations in which the BH is retained by the star, while crosses indicate cases in which the BH escapes. The rotational energy is quantified by $|E_{\rm k,rot}/E_{\rm p}|$, as defined in Section~\ref{sec:bh_retention}. The approximate scaling $|E_{\rm k,rot}/E_{\rm p}|\propto\tilde{l}^2$ at low $\tilde{l}$ and the subsequent plateau at $\gtrsim 0.1$ (\emph{blue shaded}) are evident.}
    \label{fig:rotation_supported}
\end{figure}

The orbital angular momentum from the BH is transmitted to the stellar envelope, causing the rotation of the envelope. In Figure~\ref{fig:bound_mass_radius}, we characterize the rotational support of the remnant using the ratio of rotational kinetic energy to the magnitude of the gravitational potential energy,
$|E_{\rm k,rot}/E_{\rm p}|$ (as defined in Section~\ref{sec:bh_retention}). For the $100\,M_\odot$ BH, which is too massive to retain most of the
original stellar mass to form a BH* (based on Equation~\ref{equ:critical_mass_for_bh_star}), it can 
nonetheless capture a rotationally supported 
disk-like remnant envelope, with  a largely reduced 
mass and $|E_{\rm k,rot}/E_{\rm p}|\gtrsim 0.1$.

To quantify the connection between the angular momentum of the BH and the rotational support of the remnant, we define a dimensionless angular momentum
\begin{equation}
    \tilde{l} =
    \frac{l(M_\bullet/M_\star)}
    {\sqrt{G(M_\bullet+M_\star)R_\star}},
\end{equation}
where $l$ and $M_\star$ are the specific orbital 
angular momentum of the BH and the stellar mass
prior to the collision. Figure~\ref{fig:rotation_supported} shows the relation between $\tilde{l}$ and $|E_{\rm k,rot}/E_{\rm p}|$ for simulations spanning a range of intruding BH masses and orbital angular momenta. At relatively low angular momentum, $\tilde{l}\lesssim0.5$, the rotational energy increases approximately as $|E_{\rm k,rot}/E_{\rm p}| \approx \tilde{l}^2/2$. This behavior is expected when the angular momentum of the remnant increases without substantially changing its overall structure: the rotational kinetic energy scales approximately as the square of the angular velocity, and hence increases rapidly with the injected angular momentum.

At intermediate angular momentum, $0.5\lesssim\tilde{l}\lesssim0.75$, the rotational-energy ratio instead approaches a plateau of $\sim0.15$. This indicates that the remnant responds to increasing angular momentum by becoming increasingly extended and rotationally supported. In this regime, the stellar envelope develops a disk-like configuration around the retained BH (see Figure~\ref{fig:vis_fiducial}). The approximate saturation of $|E_{\rm k,rot}/E_{\rm p}|$ therefore reflects a transition from a relatively compact, weakly rotating envelope to a configuration in which rotation provides substantial dynamical support. 

However, if $\tilde{l}\gtrsim1$, the BH 
can no longer be retained by the star.
The formation of the disk-like envelope is therefore associated with an intermediate range of the angular momentum parameter: the BH must carry sufficient angular momentum to impart substantial rotation to the stellar material, but not so much that the orbital energy and angular momentum prevent its capture.

In summary, the formation of the disk-like envelope is relevant to massive BHs with moderate orbital angular momentum, such that the stellar material is being disrupted and the orbital angular momentum is being transferred. This is the product of a micro-TDE, which can be less relevant to the main topic (BH*) of this research.  We plan to address the details of these micro-TDEs in future studies.

}

\subsection{Sequential evolution of the BH*}

While our simulations generate hydrostatic structures, the long-term evolution of these BH*s remains uncertain. Below are some speculations.

{
Since the stellar core is extremely optically thick, once the BH has inspiraled to the stellar core, the energy flux from the BH is capped by convection, at $F_{\rm conv}\sim p c_{\rm s}$ \citep[][]{BegelmanRossiArmitage_2008MNRAS.387.1649B,BallToutZytkow_2011MNRAS.414.2751B,CoughlinBegelman_2024ApJ...970..158C}.  In massive stars, the nuclear energy generation rate is self-regulated to maintain an Eddington-limited total stellar luminosity.
When the embedded BH is much less massive than its host star, its accretion luminosity can only be a small fraction of the stellar nuclear energy generation rate. Moreover, the BH's additional gravity is offset by the radiation pressure due to the release of its accretion energy.  Consequently, the global stellar structure is expected to remain largely unchanged. In contrast, a relatively massive central BH (with $M_{\rm \bullet}/M_\star \gtrsim 0.1-0.2$) may add sufficiently large energy flux to the host star's nuclear energy generation for its total luminosity to exceed the Eddington limit of the BH* \citep[][]{CoughlinBegelman_2024ApJ...970..158C}, making it thermally unstable and explode like a transient \citep[][]{HassanPernaCantiello_2026ApJ...998...65H}, particularly if an accretion disk is formed around the BH \citep[][]{GottliebCantielloNorton_2026arXiv260602700G,CantielloGottliebNorton_2026arXiv260602726C}. These uncertainties can be explored with future stellar evolution and hydrodynamical simulations.



}

A detailed calculation of the spectral energy distribution of BH*s remains beyond the scope of this work. However, similar objects have recently been proposed to explain LRDs \citep[e.g.,][]{BegelmanDexter_2026ApJ...996...48B,InayoshiMuraseKashiyama_2026ApJ..1000...90I}, although the BH*s studied here are considerably less massive. 

\subsection{Astrophysical implications}

\subsubsection{Gravitational-wave events}

Dense star clusters and AGN disks are the two leading formation channels for stellar-mass BH mergers, and both environments naturally contain main-sequence stars. Previous studies have shown that star--sBH collisions occur in both settings \citep[][]{TagawaHaimanKocsis_2020ApJ...898...25T,ChenLin_2024ApJ...967...88C,RantalaNaabLahen_2024MNRAS.531.3770R}, raising the possibility that a BH binary may merge \emph{inside} a stellar envelope.

A possible evolutionary pathway is that a stellar-mass BH is first captured by a massive star, forming a BH*. A second BH subsequently collides with the BH*, loses orbital energy through gas drag, and forms a bound binary with the embedded BH. Gas drag and gravitational torques may then accelerate the inspiral toward merger.

As an illustration, two $10\,M_\odot$ BHs in a circular orbit with semi-major axis $a=0.1\,R_\odot$ would merge through gravitational-wave emission alone in
\begin{align}
    \tau_{\rm merge}=\frac{5}{256}\frac{c^5a^4}{G^3m_1m_2(m_1+m_2)}
    \approx7.5\,{\rm yr},
\end{align}
with eccentric binaries merging considerably faster \citep{Peters_1964PhRv..136.1224P}. Gas drag is expected to shorten this timescale further \citep[][]{O'NeillD'OrazioSamsing_2024ApJ...974..216O}.

A BH merger in the stellar envelope could also produce an electromagnetic counterpart: energy deposited into the envelope by the GW recoil, or post-merger accretion onto the remnant BH, may generate optical or X-ray emission. 

\subsubsection{Dense star clusters}

{
In dense star clusters, runaway stellar collisions can lead to the formation of a supermassive star (SMS) with a mass of $\sim 10^3$--$10^5\,M_\odot$ \citep[e.g.,][]{PortegiesZwartMcMillan_2002ApJ...576..899P,KremerSperaBecker_2020ApJ...903...45K,ShiGrudicHopkins_2021MNRAS.505.2753S,RantalaNaabLahen_2024MNRAS.531.3770R}. Meanwhile, stellar-mass BHs can be repeatedly captured by the SMS and undergo hierarchical BH--BH mergers (Hu et al. submitted),
potentially building up a massive BH embedded within the SMS and thereby forming a BH*. This hierarchical merger process could grow the central BH into the IMBH regime, similar to the scenario proposed by \citet{KritosSilk_2026ApJ..1004..226K}, but potentially more efficiently because gas drag within the stellar envelope can 
damp the BHs' orbital motion on a dynamical timescale. 
A key challenge for this scenario is the retention of merger remnants against GW recoil kicks. For $M_{\rm SMS}\sim10^3$--$10^4\,M_\odot$ and $R_{\rm SMS}\sim10^3\,R_\odot$, the escape velocity is $v_{\rm esc}\sim{\rm few}\times10^2\,\rm km\,s^{-1}$, comparable to typical GW recoil velocities \citep[e.g.,][]{Holley-BockelmannGultekinShoemaker_2008ApJ...686..829H}. However, recoil velocities can be substantially reduced for unequal-mass mergers or favorable spin configurations, while gas drag within the stellar envelope may further enhance the retention of merger remnants.
}

This mechanism may also be relevant to the little red dots (LRDs) discovered by JWST at $z\gtrsim4$ \citep[e.g.,][]{MattheeNaiduBrammer_2024ApJ...963..129M}, for which BH* models have recently been proposed \citep[][]{deGraaffHvidingNaidu_2026MNRAS.551g1567D,BegelmanDexter_2026ApJ...996...48B,SunNaiduMatthee_2026OJAp....962505S,InayoshiMuraseKashiyama_2026ApJ..1000...90I}. Once the first BH* forms through a star--BH collision, subsequent runaway stellar collisions can replenish the envelope, while additional stellar-mass BHs are efficiently captured and merge with the central BH \citep[also see][]{Rantala_2026arXiv260422924R}. Quantifying the importance of this channel requires realistic models of collision rates and BH growth, which we leave for future work.

\subsubsection{AGN disks}

AGN disks provide another promising environment for BH* formation because frequent star--BH encounters are expected in their dense gaseous disks \citep[e.g.,][]{ChenLin_2024ApJ...967...88C}, while the high ambient gas density fundamentally modifies stellar evolution \citep[e.g.,][]{CantielloJermynLin_2021ApJ...910...94C,Ali-DibLin_2023MNRAS.526.5824A,XuChenLin_2026ApJ...997..206X}. Star--sBH collisions may produce BH*s in these environments, and they may also reach an accretion--wind equilibrium in the disk, potentially leading to a distinct class of stellar evolution. 

Moreover, high-velocity encounters can strongly perturb the star. Shock heating mixes the stellar interior and ejects part of the envelope into the surrounding AGN disk, contributing to the chemical enrichment. Such enrichment may help explain the anomalous abundance patterns inferred from AGN broad-line regions \citep[e.g.,][]{IsobeMaiolinoD'Eugenio_2025MNRAS.541L..71I} if the star--sBH collision rates are sufficiently high (e.g., $10^{-6}\,\rm yr^{-1}$ per star).

\section{Conclusions}
\label{sec:conclusions}

In this work, we have studied collisions between stellar-mass black holes (sBHs) and a $100\,M_\odot$ main-sequence star using both semi-analytic and hydrodynamical simulations. While the outcomes of star--sBH collisions depend on $M_\bullet/M_\star$ and other velocity parameters, we focused on the case of $M_\bullet/M_\star\lesssim 1$ and \emph{direct geometric collision}. Our main findings are as follows.

\begin{enumerate}
    \item \emph{BHs with $M_\bullet\lesssim 0.2\,M_\star$ can be retained 
    {after low-velocity collisions with stars}
    to form a BH*,} as predicted from both our semi-analytic and hydrodynamical simulations. These collisions produce a \emph{quasi-hydrostatic BH*} with a dense ($\sim 1\,\rm g\,cm^{-3}$), spherical, and extended envelope that retains most of the original stellar mass. {Our one-dimensional heating calculations indicate that no thermal runaway will be triggered during this process.} 

    \item \emph{BHs with $M_\bullet\gtrsim 0.2\,M_\star$ may not form a BH* {with substantial retained envelopes}
    through sBH--star collisions.} Above this critical mass, the shock heating released by gas drag is capable of disrupting the star. The envelope is quasi-spherical in head-on collisions and disk-like in {off-center} ones; however, in both cases the envelope is dynamically unstable, resulting in substantial mass loss and a transient event like a micro-TDE \citep[e.g.,][]{PeretsLiLombardi_2016ApJ...823..113P,KremerLombardiLu_2022ApJ...933..203K}.

    \item \emph{BH retention depends on impact velocity.} Gas drag efficiently retains the BH within the stellar envelope for impact velocities $v_\infty \lesssim 2 v_0= 2\sqrt{G(M_\star+M_\bullet)/R_\star}$. Above this threshold, the BH penetrates and escapes the star. Light BHs ($M_\bullet \lesssim 5\,M_\odot$) experience weaker drag and may escape even at $v_\infty \sim v_0$. Otherwise, the BH retention happens quickly, usually within several hundreds of hours (several tens of $\tau_{\rm dyn,\star}$).

    \item \emph{The collision drives large-scale chemical mixing.} The shock wake of the BH advects hydrogen-rich material from the envelope into the core, elevating the central hydrogen abundance from its pre-collision value (e.g., $X_{\rm H}$ rises from 0.2 to $\sim 0.3$ in our mock test). This replenishment of core hydrogen may impact the long-term evolution of the BH*. Compared with the progenitor star, the BH* envelope is substantially inflated after formation, while the central density near the BH remains comparable to that of the original star. 
\end{enumerate}

These results have several astrophysical implications. First, sBH binaries formed inside a BH* can merge via GW emission on timescales of $\sim \rm years$, producing a signal that is distinct from a vacuum BH--BH merger due to the friction from the surrounding gas or an electromagnetic counterpart. Second, in dense star clusters, repeated star--sBH collisions can migrate multiple sBHs to the center of a supermassive star, driving sequential mergers that rapidly build an increasingly massive central BH. Whether the merger product is retained depends on the competition between GW recoil kicks and the confining gas drag of the stellar envelope. This channel offers a formation pathway for IMBHs and may be relevant to the massive BHs inferred in the LRDs observed at $z \gtrsim 4$ by JWST. Third, in AGN disks, star--sBH collisions alter stellar evolution by replenishing core hydrogen and, in violent encounters, ejecting nucleosynthesis-enriched material into the disk, contributing to chemical enrichment distinct from supernovae or stellar winds.

BH*s in this work are not essentially stable in the long term, since our simulations track only the BH* formation that completes on the stellar dynamical timescale. The subsequent thermal relaxation and long-term evolution, including BH accretion, radiative feedback, and the eventual fate of the BH*, require dedicated stellar evolution modeling \citep[at least for the typical parameter space explored here;][]{HassanPernaCantiello_2026ApJ...998...65H}. In a companion paper, we study the formation of sBH binaries inside BH*s in detail. Future work should also address aspects like the observational signatures of such BH*s.


\begin{acknowledgments}
    {We thank the anonymous referee for comments that helped improve this article.}
    We thank Sizheng Ma and Hongxuan Jiang for sharing the results of their
    simulations and close interaction.  We also benefited from useful
    conversations with Long Wang and Yixian Chen.
    YS and NM acknowledge the support of the Natural Sciences and Engineering Research Council of Canada (NSERC) under funding reference No.~568580.
    QH acknowledges the support of the National Natural Science Foundation of China (grant No.~12173021 and 12133005). 
    Computations were performed on the ``Trillium'' supercomputer at the SciNet HPC Consortium. SciNet is funded by Innovation, Science and Economic Development Canada; the Digital Research Alliance of Canada; the Ontario Research Fund: Research Excellence; and the University of Toronto.
\end{acknowledgments}

%

\vspace{5mm}


\software{
    \texttt{REBOUNDx} \citep[][]{TamayoReinShi_2020MNRAS.491.2885T};
    \texttt{MESA} \citep[][]{PaxtonBildstenDotter_2011ApJS..192....3P,PaxtonCantielloArras_2013ApJS..208....4P,PaxtonMarchantSchwab_2015ApJS..220...15P,PaxtonSchwabBauer_2018ApJS..234...34P,PaxtonSmolecSchwab_2019ApJS..243...10P,JermynBauerSchwab_2023ApJS..265...15J};
    \texttt{GIZMO} \citep[][]{Hopkins_2015MNRAS.450...53H}
          }



\appendix

\begin{figure}
    \centering
    \includegraphics[width=\linewidth]{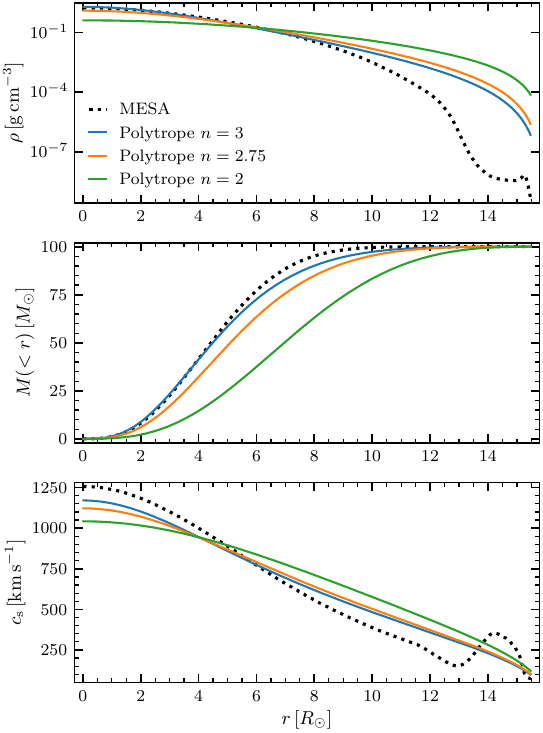}
    \caption{ Comparison of the stellar profiles adopted in this study, including the \texttt{MESA} model (cf. Section~\ref{subsec:mesa-heating}, \emph{dotted}) and polytropic models with $n=3$, $2.75$, and $2$. All stellar models have a fixed mass of $M_\star=100\,M_\odot$.}
    \label{fig:profile_comparison}
\end{figure}
\begin{figure}
    \centering
    \includegraphics[width=\linewidth]{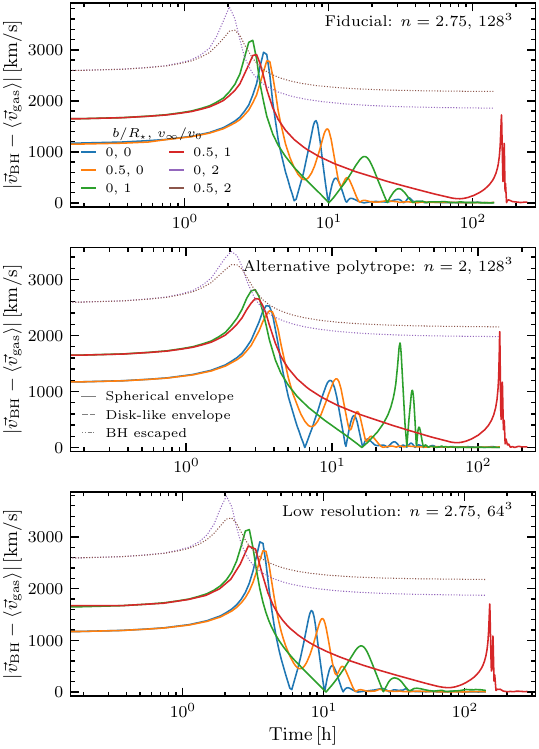}
    \caption{ Convergence tests for the hydrodynamical simulations. From top to bottom, the panels show the fiducial setup ($n=2.75$ polytrope, high resolution), an $n=2$ polytrope at high resolution, and an $n=2.75$ polytrope at low resolution. In all cases, the intruding BH has a fixed mass of $M_\bullet=10\,M_\odot$.}
    \label{fig:resolution_convergence}
\end{figure}

\section{Tests for hydrodynamical simulations}
\label{app:tests}

{
\subsection{Stellar profiles}

Figure~\ref{fig:profile_comparison} compares the stellar profiles adopted in this study, including the \texttt{MESA} model described in Section~\ref{subsec:mesa-heating} and the polytropic models used for the semi-analytic orbital integrations and hydrodynamical simulations. The \texttt{MESA} model provides a more realistic description of the internal structure of a massive star than a simple polytropic model. Nevertheless, both the density $\rho(r)$ and cumulative mass $M(<r)$ profiles show that the polytropes provide reasonable approximations to the \texttt{MESA} model over the regions containing most of the stellar mass. In particular, the $n=3$ and $n=2.75$ polytropes reproduce the central density of the \texttt{MESA} model reasonably well, with $\rho(r=0)\approx1$--$2\,\rm g\,cm^{-3}$. The $n=2$ model has a somewhat lower central density of $\sim0.4\,\rm g\,cm^{-3}$, but remains broadly similar. Near the stellar surface ($r\gtrsim10\,R_\odot$), the polytropes are substantially denser than the \texttt{MESA} model. However, most of the stellar mass is already enclosed within $r\lesssim10\,R_\odot$, and the contribution of these low-density outer layers to the hydrodynamic drag is therefore small. This is particularly relevant because $F_{\rm drag}\propto\rho$. The models also yield similar central sound speeds of $\sim1000\,\rm km\,s^{-1}$.

We adopt polytropic models for the hydrodynamical simulations because their analytic structure and equation of state make them straightforward to initialize and evolve numerically, while still providing a reasonable approximation to more detailed stellar models. Among the models considered here, the $n=3$ polytrope is physically motivated for massive stars because of their strong radiation-pressure support. However, an $n=3$ polytrope is marginally unstable to radial perturbations. We therefore adopt an $n=2.75$ polytrope as our fiducial stellar model, which provides a similar internal structure while avoiding the marginal stability of the $n=3$ case.

\subsection{Resolution convergence tests}

Figure~\ref{fig:resolution_convergence} presents tests of both the stellar profile and numerical resolution. We compare the fiducial $n=2.75$ model with an $n=2$ polytrope at the same resolution, as well as with a lower-resolution $n=2.75$ model. In the latter case, the number of resolving elements is reduced from $128^3$ to $64^3$. For both resolutions, the smoothing length is fixed at $h=0.5\,R_\odot$. We compare the evolution of the relative star--BH velocity (see also Figure~\ref{fig:bh_vel_evo}). The $n=2$ and $n=2.75$ models exhibit very similar velocity evolution, indicating that our results are not sensitive to the precise choice of polytropic index. Likewise, the high- and low-resolution $n=2.75$ simulations show nearly identical evolution, demonstrating that the fiducial results are well converged with respect to numerical resolution.

}


\bibliography{analytic,bh,star}{}

@ARTICLE{deGraaffHvidingNaidu_2026MNRAS.551g1567D,
       author = {{de Graaff}, Anna and {Hviding}, Raphael E. and {Naidu}, Rohan P. and {Greene}, Jenny E. and {Miller}, Tim B. and {Leja}, Joel and {Matthee}, Jorryt and {Brammer}, Gabriel and {Katz}, Harley and {Bezanson}, Rachel and et al.},
        title = "{Little red dots host black hole stars: a unified family of gas-reddened AGN revealed by JWST/NIRSpec spectroscopy}",
      journal = {\mnras},
         year = 2026,
        month = sep,
       volume = {551},
       number = {3},
          eid = {stag1567},
        pages = {stag1567},
          doi = {10.1093/mnras/stag1567},
archivePrefix = {arXiv},
       eprint = {2511.21820},
 primaryClass = {astro-ph.GA},
       adsurl = {https://ui.adsabs.harvard.edu/abs/2026MNRAS.551g1567D}
}

@ARTICLE{NaiduMattheeKatz_2026Natur.656..329N,
       author = {{Naidu}, Rohan P. and {Matthee}, Jorryt and {Katz}, Harley and {de Graaff}, Anna and {Oesch}, Pascal A. and {Smith}, Aaron and {Greene}, Jenny E. and {Brammer}, Gabriel and {Weibel}, Andrea and {Hviding}, Raphael and et al.},
        title = "{A gas-enshrouded and gas-reddened black hole at cosmic dawn}",
      journal = {\nat},
         year = 2026,
        month = aug,
       volume = {656},
       number = {8127},
        pages = {329-333},
          doi = {10.1038/s41586-026-10846-4},
archivePrefix = {arXiv},
       eprint = {2503.16596},
 primaryClass = {astro-ph.GA},
       adsurl = {https://ui.adsabs.harvard.edu/abs/2026Natur.656..329N}
}

@ARTICLE{KokorevChisholmNaidu_2026ApJ..1004..153K,
       author = {{Kokorev}, Vasily and {Chisholm}, John and {Naidu}, Rohan P. and {Fujimoto}, Seiji and {Atek}, Hakim and {Brammer}, Gabriel and {Finkelstein}, Steven L. and {Akins}, Hollis B. and {Berg}, Danielle A. and {Furtak}, Lukas J. and et al.},
        title = "{The Deepest GLIMPSE of a Dense Gas Cocoon Enshrouding a Little Red Dot}",
      journal = {\apj},
         year = 2026,
        month = jun,
       volume = {1004},
       number = {2},
          eid = {153},
        pages = {153},
          doi = {10.3847/1538-4357/ae4ed7},
archivePrefix = {arXiv},
       eprint = {2511.07515},
 primaryClass = {astro-ph.GA},
       adsurl = {https://ui.adsabs.harvard.edu/abs/2026ApJ..1004..153K}
}

@ARTICLE{KritosSilk_2026ApJ..1004..226K,
       author = {{Kritos}, Konstantinos and {Silk}, Joseph},
        title = "{From Nuclear Star Clusters to Little Red Dots: Black Hole Growth, Mergers, and Tidal Disruptions}",
      journal = {\apj},
         year = 2026,
        month = jun,
       volume = {1004},
       number = {2},
          eid = {226},
        pages = {226},
          doi = {10.3847/1538-4357/ae731c},
archivePrefix = {arXiv},
       eprint = {2510.21709},
 primaryClass = {astro-ph.HE},
       adsurl = {https://ui.adsabs.harvard.edu/abs/2026ApJ..1004..226K}
}

@ARTICLE{SunNaiduMatthee_2026OJAp....962505S,
       author = {{Sun}, Wendy Q. and {Naidu}, Rohan P. and {Matthee}, Jorryt and {de Graaff}, Anna and {Chisholm}, John and {Greene}, Jenny E. and {Oesch}, Pascal A. and {Torralba}, Alberto and {Hviding}, Raphael E. and {Brammer}, Gabriel and et al.},
        title = "{Little Red Dot {\ensuremath{-}} Host Galaxy = Black Hole Star: A Gas-Enshrouded Heart at the Center of Every Little Red Dot}",
      journal = {The Open Journal of Astrophysics},
         year = 2026,
        month = may,
       volume = {9},
        pages = {62505},
          doi = {10.33232/001c.162505},
archivePrefix = {arXiv},
       eprint = {2601.20929},
 primaryClass = {astro-ph.GA},
       adsurl = {https://ui.adsabs.harvard.edu/abs/2026OJAp....962505S}
}

@ARTICLE{Rantala_2026arXiv260422924R,
       author = {{Rantala}, Antti},
        title = "{Supermassive stars with embedded stellar black hole cores: dense assembling star clusters as faint multiple Little Red Dot systems}",
      journal = {arXiv e-prints},
         year = 2026,
        month = apr,
          eid = {arXiv:2604.22924},
        pages = {arXiv:2604.22924},
          doi = {10.48550/arXiv.2604.22924},
archivePrefix = {arXiv},
       eprint = {2604.22924},
 primaryClass = {astro-ph.GA},
       adsurl = {https://ui.adsabs.harvard.edu/abs/2026arXiv260422924R}
}

@ARTICLE{RastelloIorioGieles_2026A&A...707A.217R,
       author = {{Rastello}, Sara and {Iorio}, Giuliano and {Gieles}, Mark and {Wang}, Long},
        title = "{Micro-tidal disruption events in young star clusters}",
      journal = {\aap},
         year = 2026,
        month = mar,
       volume = {707},
          eid = {A217},
        pages = {A217},
          doi = {10.1051/0004-6361/202556781},
archivePrefix = {arXiv},
       eprint = {2509.07067},
 primaryClass = {astro-ph.HE},
       adsurl = {https://ui.adsabs.harvard.edu/abs/2026A&A...707A.217R}
}

@ARTICLE{InayoshiMuraseKashiyama_2026ApJ..1000...90I,
       author = {{Inayoshi}, Kohei and {Murase}, Kohta and {Kashiyama}, Kazumi},
        title = "{Spectral Uniformity of Little Red Dots: A Natural Outcome of Coevolving Seed Black Holes and Nascent Starbursts}",
      journal = {\apj},
         year = 2026,
        month = mar,
       volume = {1000},
       number = {1},
          eid = {90},
        pages = {90},
          doi = {10.3847/1538-4357/ae42ce},
archivePrefix = {arXiv},
       eprint = {2509.19422},
 primaryClass = {astro-ph.GA},
       adsurl = {https://ui.adsabs.harvard.edu/abs/2026ApJ..1000...90I}
}

@ARTICLE{BegelmanDexter_2026ApJ...996...48B,
       author = {{Begelman}, Mitchell C. and {Dexter}, Jason},
        title = "{Little Red Dots as Late-stage Quasi-stars}",
      journal = {\apj},
         year = 2026,
        month = jan,
       volume = {996},
       number = {1},
          eid = {48},
        pages = {48},
          doi = {10.3847/1538-4357/ae274a},
archivePrefix = {arXiv},
       eprint = {2507.09085},
 primaryClass = {astro-ph.GA},
       adsurl = {https://ui.adsabs.harvard.edu/abs/2026ApJ...996...48B}
}

@ARTICLE{YanagisawaOuchiGolubchik_2026arXiv260106015Y,
       author = {{Yanagisawa}, Hiroto and {Ouchi}, Masami and {Golubchik}, Miriam and {Oguri}, Masamune and {Fujimoto}, Seiji and {Kokorev}, Vasily and {Brammer}, Gabriel and {Sun}, Fengwu and {Nakane}, Minami and {Harikane}, Yuichi and et al.},
        title = "{VENUS: Two Faint Little Red Dots Separated by $\sim70\,\mathrm{pc}$ Hidden in a Single Lensed Galaxy at $z\sim7$}",
      journal = {arXiv e-prints},
         year = 2026,
        month = jan,
          eid = {arXiv:2601.06015},
        pages = {arXiv:2601.06015},
          doi = {10.48550/arXiv.2601.06015},
archivePrefix = {arXiv},
       eprint = {2601.06015},
 primaryClass = {astro-ph.GA},
       adsurl = {https://ui.adsabs.harvard.edu/abs/2026arXiv260106015Y}
}

@ARTICLE{InayoshiHo_2025arXiv251203130I,
       author = {{Inayoshi}, Kohei and {Ho}, Luis C.},
        title = "{A Critical Evaluation of the Physical Nature of the Little Red Dots}",
      journal = {arXiv e-prints},
         year = 2025,
        month = dec,
          eid = {arXiv:2512.03130},
        pages = {arXiv:2512.03130},
          doi = {10.48550/arXiv.2512.03130},
archivePrefix = {arXiv},
       eprint = {2512.03130},
 primaryClass = {astro-ph.GA},
       adsurl = {https://ui.adsabs.harvard.edu/abs/2025arXiv251203130I}
}

@ARTICLE{deGraaffRixNaidu_2025A&A...701A.168D,
       author = {{de Graaff}, Anna and {Rix}, Hans-Walter and {Naidu}, Rohan P. and {Labb{\'e}}, Ivo and {Wang}, Bingjie and {Leja}, Joel and {Matthee}, Jorryt and {Katz}, Harley and {Greene}, Jenny E. and {Hviding}, Raphael E. and et al.},
        title = "{A remarkable ruby: Absorption in dense gas, rather than evolved stars, drives the extreme Balmer break of a little red dot at z = 3.5}",
      journal = {\aap},
         year = 2025,
        month = sep,
       volume = {701},
          eid = {A168},
        pages = {A168},
          doi = {10.1051/0004-6361/202554681},
archivePrefix = {arXiv},
       eprint = {2503.16600},
 primaryClass = {astro-ph.GA},
       adsurl = {https://ui.adsabs.harvard.edu/abs/2025A&A...701A.168D}
}

@ARTICLE{KocevskiFinkelsteinBarro_2025ApJ...986..126K,
       author = {{Kocevski}, Dale D. and {Finkelstein}, Steven L. and {Barro}, Guillermo and {Taylor}, Anthony J. and {Calabr{\`o}}, Antonello and {Laloux}, Brivael and {Buchner}, Johannes and {Trump}, Jonathan R. and {Leung}, Gene C.~K. and {Yang}, Guang and et al.},
        title = "{The Rise of Faint, Red Active Galactic Nuclei at z > 4: A Sample of Little Red Dots in the JWST Extragalactic Legacy Fields}",
      journal = {\apj},
         year = 2025,
        month = jun,
       volume = {986},
       number = {2},
          eid = {126},
        pages = {126},
          doi = {10.3847/1538-4357/adbc7d},
archivePrefix = {arXiv},
       eprint = {2404.03576},
 primaryClass = {astro-ph.GA},
       adsurl = {https://ui.adsabs.harvard.edu/abs/2025ApJ...986..126K}
}

@ARTICLE{InayoshiMaiolino_2025ApJ...980L..27I,
       author = {{Inayoshi}, Kohei and {Maiolino}, Roberto},
        title = "{Extremely Dense Gas around Little Red Dots and High-redshift Active Galactic Nuclei: A Nonstellar Origin of the Balmer Break and Absorption Features}",
      journal = {\apjl},
         year = 2025,
        month = feb,
       volume = {980},
       number = {2},
          eid = {L27},
        pages = {L27},
          doi = {10.3847/2041-8213/adaebd},
archivePrefix = {arXiv},
       eprint = {2409.07805},
 primaryClass = {astro-ph.GA},
       adsurl = {https://ui.adsabs.harvard.edu/abs/2025ApJ...980L..27I}
}

@ARTICLE{KirogluKremerBiscoveanu_2025ApJ...979..237K,
       author = {{K{\i}ro{\u{g}}lu}, Fulya and {Kremer}, Kyle and {Biscoveanu}, Sylvia and {Gonz{\'a}lez Prieto}, Elena and {Rasio}, Frederic A.},
        title = "{Black Hole Accretion and Spin-up through Stellar Collisions in Dense Star Clusters}",
      journal = {\apj},
         year = 2025,
        month = feb,
       volume = {979},
       number = {2},
          eid = {237},
        pages = {237},
          doi = {10.3847/1538-4357/ada26b},
archivePrefix = {arXiv},
       eprint = {2410.01879},
 primaryClass = {astro-ph.HE},
       adsurl = {https://ui.adsabs.harvard.edu/abs/2025ApJ...979..237K}
}

@ARTICLE{RantalaNaabLahen_2024MNRAS.531.3770R,
       author = {{Rantala}, Antti and {Naab}, Thorsten and {Lah{\'e}n}, Natalia},
        title = "{FROST-CLUSTERS - I. Hierarchical star cluster assembly boosts intermediate-mass black hole formation}",
      journal = {\mnras},
         year = 2024,
        month = jul,
       volume = {531},
       number = {3},
        pages = {3770-3799},
          doi = {10.1093/mnras/stae1413},
archivePrefix = {arXiv},
       eprint = {2403.10602},
 primaryClass = {astro-ph.GA},
       adsurl = {https://ui.adsabs.harvard.edu/abs/2024MNRAS.531.3770R}
}

@ARTICLE{KokorevCaputiGreene_2024ApJ...968...38K,
       author = {{Kokorev}, Vasily and {Caputi}, Karina I. and {Greene}, Jenny E. and {Dayal}, Pratika and {Trebitsch}, Maxime and {Cutler}, Sam E. and {Fujimoto}, Seiji and {Labb{\'e}}, Ivo and {Miller}, Tim B. and {Iani}, Edoardo and et al.},
        title = "{A Census of Photometrically Selected Little Red Dots at 4 < z < 9 in JWST Blank Fields}",
      journal = {\apj},
         year = 2024,
        month = jun,
       volume = {968},
       number = {1},
          eid = {38},
        pages = {38},
          doi = {10.3847/1538-4357/ad4265},
archivePrefix = {arXiv},
       eprint = {2401.09981},
 primaryClass = {astro-ph.GA},
       adsurl = {https://ui.adsabs.harvard.edu/abs/2024ApJ...968...38K}
}

@ARTICLE{VynatheyaRyuPakmor_2024A&A...685A..45V,
       author = {{Vynatheya}, Pavan and {Ryu}, Taeho and {Pakmor}, R{\"u}diger and {de Mink}, Selma E. and {Perets}, Hagai B.},
        title = "{Simulating the tidal disruption of stars by stellar-mass black holes using moving-mesh hydrodynamics}",
      journal = {\aap},
         year = 2024,
        month = may,
       volume = {685},
          eid = {A45},
        pages = {A45},
          doi = {10.1051/0004-6361/202348357},
archivePrefix = {arXiv},
       eprint = {2310.14852},
 primaryClass = {astro-ph.HE},
       adsurl = {https://ui.adsabs.harvard.edu/abs/2024A&A...685A..45V}
}

@ARTICLE{SuzuguchiSugimuraHosokawa_2024ApJ...966....7S,
       author = {{Suzuguchi}, Tomoya and {Sugimura}, Kazuyuki and {Hosokawa}, Takashi and {Matsumoto}, Tomoaki},
        title = "{Gas Dynamical Friction on Accreting Objects}",
      journal = {\apj},
         year = 2024,
        month = may,
       volume = {966},
       number = {1},
          eid = {7},
        pages = {7},
          doi = {10.3847/1538-4357/ad34af},
archivePrefix = {arXiv},
       eprint = {2401.13032},
 primaryClass = {astro-ph.GA},
       adsurl = {https://ui.adsabs.harvard.edu/abs/2024ApJ...966....7S}
}

@ARTICLE{MattheeNaiduBrammer_2024ApJ...963..129M,
       author = {{Matthee}, Jorryt and {Naidu}, Rohan P. and {Brammer}, Gabriel and {Chisholm}, John and {Eilers}, Anna-Christina and {Goulding}, Andy and {Greene}, Jenny and {Kashino}, Daichi and {Labbe}, Ivo and {Lilly}, Simon J. and et al.},
        title = "{Little Red Dots: An Abundant Population of Faint Active Galactic Nuclei at z {\ensuremath{\sim}} 5 Revealed by the EIGER and FRESCO JWST Surveys}",
      journal = {\apj},
         year = 2024,
        month = mar,
       volume = {963},
       number = {2},
          eid = {129},
        pages = {129},
          doi = {10.3847/1538-4357/ad2345},
archivePrefix = {arXiv},
       eprint = {2306.05448},
 primaryClass = {astro-ph.GA},
       adsurl = {https://ui.adsabs.harvard.edu/abs/2024ApJ...963..129M}
}

@ARTICLE{XinHaimanPerna_2024ApJ...961..149X,
       author = {{Xin}, Chengcheng and {Haiman}, Zolt{\'a}n and {Perna}, Rosalba and {Wang}, Yihan and {Ryu}, Taeho},
        title = "{``Tidal Peeling Events'': Low-eccentricity Tidal Disruption of a Star by a Stellar-mass Black Hole}",
      journal = {\apj},
         year = 2024,
        month = feb,
       volume = {961},
       number = {2},
          eid = {149},
        pages = {149},
          doi = {10.3847/1538-4357/ad11d3},
archivePrefix = {arXiv},
       eprint = {2303.12846},
 primaryClass = {astro-ph.HE},
       adsurl = {https://ui.adsabs.harvard.edu/abs/2024ApJ...961..149X}
}

@ARTICLE{HopkinsNadlerGrudic_2023MNRAS.525.5951H,
       author = {{Hopkins}, Philip F. and {Nadler}, Ethan O. and {Grudi{\'c}}, Michael Y. and {Shen}, Xuejian and {Sands}, Isabel and {Jiang}, Fangzhou},
        title = "{Novel conservative methods for adaptive force softening in collisionless and multispecies N-body simulations}",
      journal = {\mnras},
         year = 2023,
        month = nov,
       volume = {525},
       number = {4},
        pages = {5951-5977},
          doi = {10.1093/mnras/stad2548},
archivePrefix = {arXiv},
       eprint = {2212.06851},
 primaryClass = {astro-ph.GA},
       adsurl = {https://ui.adsabs.harvard.edu/abs/2023MNRAS.525.5951H}
}

@ARTICLE{LescaudronDuboisBeckmann_2023A&A...674A.217L,
       author = {{Lescaudron}, Sandrine and {Dubois}, Yohan and {Beckmann}, Ricarda S. and {Volonteri}, Marta},
        title = "{Dynamical friction of a massive black hole in a turbulent gaseous medium}",
      journal = {\aap},
         year = 2023,
        month = jun,
       volume = {674},
          eid = {A217},
        pages = {A217},
          doi = {10.1051/0004-6361/202243392},
archivePrefix = {arXiv},
       eprint = {2209.13548},
 primaryClass = {astro-ph.GA},
       adsurl = {https://ui.adsabs.harvard.edu/abs/2023A&A...674A.217L}
}

@ARTICLE{RyuPernaWang_2022MNRAS.516.2204R,
       author = {{Ryu}, Taeho and {Perna}, Rosalba and {Wang}, Yi-Han},
        title = "{Close encounters of stars with stellar-mass black hole binaries}",
      journal = {\mnras},
         year = 2022,
        month = oct,
       volume = {516},
       number = {2},
        pages = {2204-2217},
          doi = {10.1093/mnras/stac2316},
archivePrefix = {arXiv},
       eprint = {2206.00603},
 primaryClass = {astro-ph.HE},
       adsurl = {https://ui.adsabs.harvard.edu/abs/2022MNRAS.516.2204R}
}

@ARTICLE{YangBartosFragione_2022ApJ...933L..28Y,
       author = {{Yang}, Y. and {Bartos}, I. and {Fragione}, G. and {Haiman}, Z. and {Kowalski}, M. and {M{\'a}rka}, S. and {Perna}, R. and {Tagawa}, H.},
        title = "{Tidal Disruption on Stellar-mass Black Holes in Active Galactic Nuclei}",
      journal = {\apjl},
         year = 2022,
        month = jul,
       volume = {933},
       number = {2},
          eid = {L28},
        pages = {L28},
          doi = {10.3847/2041-8213/ac7c0b},
archivePrefix = {arXiv},
       eprint = {2105.02342},
 primaryClass = {astro-ph.HE},
       adsurl = {https://ui.adsabs.harvard.edu/abs/2022ApJ...933L..28Y}
}

@ARTICLE{KremerLombardiLu_2022ApJ...933..203K,
       author = {{Kremer}, Kyle and {Lombardi}, James C. and {Lu}, Wenbin and {Piro}, Anthony L. and {Rasio}, Frederic A.},
        title = "{Hydrodynamics of Collisions and Close Encounters between Stellar Black Holes and Main-sequence Stars}",
      journal = {\apj},
         year = 2022,
        month = jul,
       volume = {933},
       number = {2},
          eid = {203},
        pages = {203},
          doi = {10.3847/1538-4357/ac714f},
archivePrefix = {arXiv},
       eprint = {2201.12368},
 primaryClass = {astro-ph.HE},
       adsurl = {https://ui.adsabs.harvard.edu/abs/2022ApJ...933..203K}
}

@ARTICLE{RoseNaozSari_2022ApJ...929L..22R,
       author = {{Rose}, Sanaea C. and {Naoz}, Smadar and {Sari}, Re'em and {Linial}, Itai},
        title = "{The Formation of Intermediate-mass Black Holes in Galactic Nuclei}",
      journal = {\apjl},
         year = 2022,
        month = apr,
       volume = {929},
       number = {2},
          eid = {L22},
        pages = {L22},
          doi = {10.3847/2041-8213/ac6426},
archivePrefix = {arXiv},
       eprint = {2201.00022},
 primaryClass = {astro-ph.GA},
       adsurl = {https://ui.adsabs.harvard.edu/abs/2022ApJ...929L..22R}
}

@ARTICLE{ShiGrudicHopkins_2021MNRAS.505.2753S,
       author = {{Shi}, Yanlong and {Grudi{\'c}}, Michael Y. and {Hopkins}, Philip F.},
        title = "{The mass budget for intermediate-mass black holes in dense star clusters}",
      journal = {\mnras},
         year = 2021,
        month = aug,
       volume = {505},
       number = {2},
        pages = {2753-2763},
          doi = {10.1093/mnras/stab1470},
archivePrefix = {arXiv},
       eprint = {2008.12290},
 primaryClass = {astro-ph.GA},
       adsurl = {https://ui.adsabs.harvard.edu/abs/2021MNRAS.505.2753S}
}

@ARTICLE{KremerSperaBecker_2020ApJ...903...45K,
       author = {{Kremer}, Kyle and {Spera}, Mario and {Becker}, Devin and {Chatterjee}, Sourav and {Di Carlo}, Ugo N. and {Fragione}, Giacomo and {Rodriguez}, Carl L. and {Ye}, Claire S. and {Rasio}, Frederic A.},
        title = "{Populating the Upper Black Hole Mass Gap through Stellar Collisions in Young Star Clusters}",
      journal = {\apj},
         year = 2020,
        month = nov,
       volume = {903},
       number = {1},
          eid = {45},
        pages = {45},
          doi = {10.3847/1538-4357/abb945},
archivePrefix = {arXiv},
       eprint = {2006.10771},
 primaryClass = {astro-ph.HE},
       adsurl = {https://ui.adsabs.harvard.edu/abs/2020ApJ...903...45K}
}

@ARTICLE{TagawaHaimanKocsis_2020ApJ...898...25T,
       author = {{Tagawa}, Hiromichi and {Haiman}, Zolt{\'a}n and {Kocsis}, Bence},
        title = "{Formation and Evolution of Compact-object Binaries in AGN Disks}",
      journal = {\apj},
         year = 2020,
        month = jul,
       volume = {898},
       number = {1},
          eid = {25},
        pages = {25},
          doi = {10.3847/1538-4357/ab9b8c},
archivePrefix = {arXiv},
       eprint = {1912.08218},
 primaryClass = {astro-ph.GA},
       adsurl = {https://ui.adsabs.harvard.edu/abs/2020ApJ...898...25T}
}

@ARTICLE{KremerLuRodriguez_2019ApJ...881...75K,
       author = {{Kremer}, Kyle and {Lu}, Wenbin and {Rodriguez}, Carl L. and {Lachat}, Mitchell and {Rasio}, Frederic A.},
        title = "{Tidal Disruptions of Stars by Black Hole Remnants in Dense Star Clusters}",
      journal = {\apj},
         year = 2019,
        month = aug,
       volume = {881},
       number = {1},
          eid = {75},
        pages = {75},
          doi = {10.3847/1538-4357/ab2e0c},
archivePrefix = {arXiv},
       eprint = {1904.06353},
 primaryClass = {astro-ph.HE},
       adsurl = {https://ui.adsabs.harvard.edu/abs/2019ApJ...881...75K}
}

@ARTICLE{PeretsLiLombardi_2016ApJ...823..113P,
       author = {{Perets}, Hagai B. and {Li}, Zhuo and {Lombardi}, Jr., James C. and {Milcarek}, Jr., Stephen R.},
        title = "{Micro-tidal Disruption Events by Stellar Compact Objects and the Production of Ultra-long GRBs}",
      journal = {\apj},
         year = 2016,
        month = jun,
       volume = {823},
       number = {2},
          eid = {113},
        pages = {113},
          doi = {10.3847/0004-637X/823/2/113},
archivePrefix = {arXiv},
       eprint = {1602.07698},
 primaryClass = {astro-ph.HE},
       adsurl = {https://ui.adsabs.harvard.edu/abs/2016ApJ...823..113P}
}

@ARTICLE{Hopkins_2015MNRAS.450...53H,
       author = {{Hopkins}, Philip F.},
        title = "{A new class of accurate, mesh-free hydrodynamic simulation methods}",
      journal = {\mnras},
         year = 2015,
        month = jun,
       volume = {450},
       number = {1},
        pages = {53-110},
          doi = {10.1093/mnras/stv195},
archivePrefix = {arXiv},
       eprint = {1409.7395},
 primaryClass = {astro-ph.CO},
       adsurl = {https://ui.adsabs.harvard.edu/abs/2015MNRAS.450...53H}
}

@ARTICLE{VolonteriBegelman_2010MNRAS.409.1022V,
       author = {{Volonteri}, Marta and {Begelman}, Mitchell C.},
        title = "{Quasi-stars and the cosmic evolution of massive black holes}",
      journal = {\mnras},
         year = 2010,
        month = dec,
       volume = {409},
       number = {3},
        pages = {1022-1032},
          doi = {10.1111/j.1365-2966.2010.17359.x},
archivePrefix = {arXiv},
       eprint = {1003.5220},
 primaryClass = {astro-ph.HE},
       adsurl = {https://ui.adsabs.harvard.edu/abs/2010MNRAS.409.1022V}
}

@ARTICLE{Holley-BockelmannGultekinShoemaker_2008ApJ...686..829H,
       author = {{Holley-Bockelmann}, Kelly and {G{\"u}ltekin}, Kayhan and {Shoemaker}, Deirdre and {Yunes}, Nicolas},
        title = "{Gravitational Wave Recoil and the Retention of Intermediate-Mass Black Holes}",
      journal = {\apj},
         year = 2008,
        month = oct,
       volume = {686},
       number = {2},
        pages = {829-837},
          doi = {10.1086/591218},
archivePrefix = {arXiv},
       eprint = {0707.1334},
 primaryClass = {astro-ph},
       adsurl = {https://ui.adsabs.harvard.edu/abs/2008ApJ...686..829H}
}

@ARTICLE{BegelmanRossiArmitage_2008MNRAS.387.1649B,
       author = {{Begelman}, Mitchell C. and {Rossi}, Elena M. and {Armitage}, Philip J.},
        title = "{Quasi-stars: accreting black holes inside massive envelopes}",
      journal = {\mnras},
         year = 2008,
        month = jul,
       volume = {387},
       number = {4},
        pages = {1649-1659},
          doi = {10.1111/j.1365-2966.2008.13344.x},
archivePrefix = {arXiv},
       eprint = {0711.4078},
 primaryClass = {astro-ph},
       adsurl = {https://ui.adsabs.harvard.edu/abs/2008MNRAS.387.1649B}
}

@BOOK{BinneyTremaine_2008gady.book.....B,
       author = {{Binney}, James and {Tremaine}, Scott},
        title = "{Galactic Dynamics: Second Edition}",
         year = 2008,
       adsurl = {https://ui.adsabs.harvard.edu/abs/2008gady.book.....B}
}

@ARTICLE{Edgar_2004NewAR..48..843E,
       author = {{Edgar}, Richard},
        title = "{A review of Bondi-Hoyle-Lyttleton accretion}",
      journal = {\nar},
         year = 2004,
        month = sep,
       volume = {48},
       number = {10},
        pages = {843-859},
          doi = {10.1016/j.newar.2004.06.001},
archivePrefix = {arXiv},
       eprint = {astro-ph/0406166},
 primaryClass = {astro-ph},
       adsurl = {https://ui.adsabs.harvard.edu/abs/2004NewAR..48..843E}
}

@ARTICLE{PortegiesZwartMcMillan_2002ApJ...576..899P,
       author = {{Portegies Zwart}, Simon F. and {McMillan}, Stephen L.~W.},
        title = "{The Runaway Growth of Intermediate-Mass Black Holes in Dense Star Clusters}",
      journal = {\apj},
         year = 2002,
        month = sep,
       volume = {576},
       number = {2},
        pages = {899-907},
          doi = {10.1086/341798},
archivePrefix = {arXiv},
       eprint = {astro-ph/0201055},
 primaryClass = {astro-ph},
       adsurl = {https://ui.adsabs.harvard.edu/abs/2002ApJ...576..899P}
}

@ARTICLE{Ostriker_1999ApJ...513..252O,
       author = {{Ostriker}, Eve C.},
        title = "{Dynamical Friction in a Gaseous Medium}",
      journal = {\apj},
         year = 1999,
        month = mar,
       volume = {513},
       number = {1},
        pages = {252-258},
          doi = {10.1086/306858},
archivePrefix = {arXiv},
       eprint = {astro-ph/9810324},
 primaryClass = {astro-ph},
       adsurl = {https://ui.adsabs.harvard.edu/abs/1999ApJ...513..252O}
}

@ARTICLE{ToutPolsEggleton_1996MNRAS.281..257T,
       author = {{Tout}, Christopher A. and {Pols}, Onno R. and {Eggleton}, Peter P. and {Han}, Zhanwen},
        title = "{Zero-age main-seqence radii and luminosities as analytic functions of mass and metallicity}",
      journal = {\mnras},
         year = 1996,
        month = jul,
       volume = {281},
       number = {1},
        pages = {257-262},
          doi = {10.1093/mnras/281.1.257},
       adsurl = {https://ui.adsabs.harvard.edu/abs/1996MNRAS.281..257T}
}

@ARTICLE{ThorneZytkow_1977ApJ...212..832T,
       author = {{Thorne}, K.~S. and {Zytkow}, A.~N.},
        title = "{Stars with degenerate neutron cores. I. Structure of equilibrium models.}",
      journal = {\apj},
         year = 1977,
        month = mar,
       volume = {212},
        pages = {832-858},
          doi = {10.1086/155109},
       adsurl = {https://ui.adsabs.harvard.edu/abs/1977ApJ...212..832T}
}

@ARTICLE{Peters_1964PhRv..136.1224P,
       author = {{Peters}, P.~C.},
        title = "{Gravitational Radiation and the Motion of Two Point Masses}",
      journal = {Physical Review},
         year = 1964,
        month = nov,
       volume = {136},
       number = {4B},
        pages = {1224-1232},
          doi = {10.1103/PhysRev.136.B1224},
       adsurl = {https://ui.adsabs.harvard.edu/abs/1964PhRv..136.1224P}
}

@ARTICLE{Bondi_1952MNRAS.112..195B,
       author = {{Bondi}, H.},
        title = "{On spherically symmetrical accretion}",
      journal = {\mnras},
         year = 1952,
        month = jan,
       volume = {112},
        pages = {195},
          doi = {10.1093/mnras/112.2.195},
       adsurl = {https://ui.adsabs.harvard.edu/abs/1952MNRAS.112..195B}
}

@ARTICLE{Chandrasekhar_1943ApJ....97..255C,
       author = {{Chandrasekhar}, S.},
        title = "{Dynamical Friction. I. General Considerations: the Coefficient of Dynamical Friction.}",
      journal = {\apj},
         year = 1943,
        month = mar,
       volume = {97},
        pages = {255},
          doi = {10.1086/144517},
       adsurl = {https://ui.adsabs.harvard.edu/abs/1943ApJ....97..255C}
}

@ARTICLE{HoyleLyttleton_1939PCPS...35..405H,
       author = {{Hoyle}, F. and {Lyttleton}, R.~A.},
        title = "{The effect of interstellar matter on climatic variation}",
      journal = {Proceedings of the Cambridge Philosophical Society},
         year = 1939,
        month = jan,
       volume = {35},
       number = {3},
        pages = {405},
          doi = {10.1017/S0305004100021150},
       adsurl = {https://ui.adsabs.harvard.edu/abs/1939PCPS...35..405H}
}

@ARTICLE{ShiHuangLin_2026arXiv260800242S,
       author = {{Shi}, Yanlong and {Huang}, Xiaoshan and {Lin}, Douglas N.~C. and {Murray}, Norman},
        title = "{Stellar mergers and chemical element mixing: implications for the metamorphic stellar evolution in AGN disks}",
      journal = {arXiv e-prints},
         year = 2026,
        month = jul,
          eid = {arXiv:2608.00242},
        pages = {arXiv:2608.00242},
          doi = {10.48550/arXiv.2608.00242},
archivePrefix = {arXiv},
       eprint = {2608.00242},
 primaryClass = {astro-ph.GA},
       adsurl = {https://ui.adsabs.harvard.edu/abs/2026arXiv260800242S}
}

@ARTICLE{CantielloGottliebNorton_2026arXiv260602726C,
       author = {{Cantiello}, Matteo and {Gottlieb}, Ore and {Norton}, Cameron and {Kleban}, Matthew and {Van Tilburg}, Ken},
        title = "{Accretion of Primordial Black Holes in Stellar Interiors}",
      journal = {arXiv e-prints},
         year = 2026,
        month = jun,
          eid = {arXiv:2606.02726},
        pages = {arXiv:2606.02726},
          doi = {10.48550/arXiv.2606.02726},
archivePrefix = {arXiv},
       eprint = {2606.02726},
 primaryClass = {astro-ph.HE},
       adsurl = {https://ui.adsabs.harvard.edu/abs/2026arXiv260602726C}
}

@ARTICLE{GottliebCantielloNorton_2026arXiv260602700G,
       author = {{Gottlieb}, Ore and {Cantiello}, Matteo and {Norton}, Cameron and {Van Tilburg}, Ken and {Kleban}, Matthew},
        title = "{The Life and Death of Stars That Capture Primordial Black Holes}",
      journal = {arXiv e-prints},
         year = 2026,
        month = jun,
          eid = {arXiv:2606.02700},
        pages = {arXiv:2606.02700},
          doi = {10.48550/arXiv.2606.02700},
archivePrefix = {arXiv},
       eprint = {2606.02700},
 primaryClass = {astro-ph.HE},
       adsurl = {https://ui.adsabs.harvard.edu/abs/2026arXiv260602700G}
}

@ARTICLE{HassanPernaCantiello_2026ApJ...998...65H,
       author = {{Hassan}, Jake B. and {Perna}, Rosalba and {Cantiello}, Matteo and {Armitage}, Philip J. and {Begelman}, Mitchell C. and {Ryu}, Taeho},
        title = "{The Growth of the Central Black Holes in Quasi-stars}",
      journal = {\apj},
         year = 2026,
        month = feb,
       volume = {998},
       number = {1},
          eid = {65},
        pages = {65},
          doi = {10.3847/1538-4357/ae3002},
archivePrefix = {arXiv},
       eprint = {2510.18301},
 primaryClass = {astro-ph.SR},
       adsurl = {https://ui.adsabs.harvard.edu/abs/2026ApJ...998...65H}
}

@ARTICLE{XuChenLin_2026ApJ...997..206X,
       author = {{Xu}, Zheng-Hao and {Chen}, Yi-Xian and {Lin}, Douglas N.~C.},
        title = "{Stellar Evolution with Radiative Feedback in AGN Disks}",
      journal = {\apj},
         year = 2026,
        month = feb,
       volume = {997},
       number = {2},
          eid = {206},
        pages = {206},
          doi = {10.3847/1538-4357/ae2271},
archivePrefix = {arXiv},
       eprint = {2511.03904},
 primaryClass = {astro-ph.GA},
       adsurl = {https://ui.adsabs.harvard.edu/abs/2026ApJ...997..206X}
}

@ARTICLE{Xu_2025RAA....25k5013X,
       author = {{Xu}, Zheng-Hao},
        title = "{The Fate of Stars Embedded in AGN Disks is Determined by an Internal Mixing Threshold}",
      journal = {Research in Astronomy and Astrophysics},
         year = 2025,
        month = nov,
       volume = {25},
       number = {11},
          eid = {115013},
        pages = {115013},
          doi = {10.1088/1674-4527/adfeb9},
       adsurl = {https://ui.adsabs.harvard.edu/abs/2025RAA....25k5013X}
}

@ARTICLE{TsunaLu_2025ApJ...986...84T,
       author = {{Tsuna}, Daichi and {Lu}, Wenbin},
        title = "{Stellar Tidal Disruptions by Newborn Neutron Stars or Black Holes: A Mechanism for Hydrogen-poor (Super)luminous Supernovae and Fast Blue Optical Transients}",
      journal = {\apj},
         year = 2025,
        month = jun,
       volume = {986},
       number = {1},
          eid = {84},
        pages = {84},
          doi = {10.3847/1538-4357/add158},
archivePrefix = {arXiv},
       eprint = {2501.03316},
 primaryClass = {astro-ph.HE},
       adsurl = {https://ui.adsabs.harvard.edu/abs/2025ApJ...986...84T}
}

@ARTICLE{CoughlinBegelman_2024ApJ...970..158C,
       author = {{Coughlin}, Eric R. and {Begelman}, Mitchell C.},
        title = "{Quasi-stars as a Means of Rapid Black Hole Growth in the Early Universe}",
      journal = {\apj},
         year = 2024,
        month = aug,
       volume = {970},
       number = {2},
          eid = {158},
        pages = {158},
          doi = {10.3847/1538-4357/ad5723},
archivePrefix = {arXiv},
       eprint = {2405.00084},
 primaryClass = {astro-ph.GA},
       adsurl = {https://ui.adsabs.harvard.edu/abs/2024ApJ...970..158C}
}

@ARTICLE{ChenLin_2024ApJ...967...88C,
       author = {{Chen}, Yi-Xian and {Lin}, Douglas N.~C.},
        title = "{The Population of Massive Stars in Active Galactic Nuclei Disks}",
      journal = {\apj},
         year = 2024,
        month = jun,
       volume = {967},
       number = {2},
          eid = {88},
        pages = {88},
          doi = {10.3847/1538-4357/ad3c3a},
archivePrefix = {arXiv},
       eprint = {2404.08780},
 primaryClass = {astro-ph.GA},
       adsurl = {https://ui.adsabs.harvard.edu/abs/2024ApJ...967...88C}
}

@ARTICLE{Ali-DibLin_2023MNRAS.526.5824A,
       author = {{Ali-Dib}, Mohamad and {Lin}, Douglas N.~C.},
        title = "{The impermanent fate of massive stars in AGN discs}",
      journal = {\mnras},
         year = 2023,
        month = dec,
       volume = {526},
       number = {4},
        pages = {5824-5838},
          doi = {10.1093/mnras/stad2774},
archivePrefix = {arXiv},
       eprint = {2309.04392},
 primaryClass = {astro-ph.GA},
       adsurl = {https://ui.adsabs.harvard.edu/abs/2023MNRAS.526.5824A}
}

@ARTICLE{RopkeDeMarco_2023LRCA....9....2R,
       author = {{R{\"o}pke}, Friedrich K. and {De Marco}, Orsola},
        title = "{Simulations of common-envelope evolution in binary stellar systems: physical models and numerical techniques}",
      journal = {Living Reviews in Computational Astrophysics},
         year = 2023,
        month = dec,
       volume = {9},
       number = {1},
          eid = {2},
        pages = {2},
          doi = {10.1007/s41115-023-00017-x},
archivePrefix = {arXiv},
       eprint = {2212.07308},
 primaryClass = {astro-ph.SR},
       adsurl = {https://ui.adsabs.harvard.edu/abs/2023LRCA....9....2R}
}

@ARTICLE{GenerozovPerets_2023MNRAS.522.1763G,
       author = {{Generozov}, A. and {Perets}, H.~B.},
        title = "{Capture of stars into gaseous discs around massive black holes: alignment, circularization, and growth}",
      journal = {\mnras},
         year = 2023,
        month = jun,
       volume = {522},
       number = {2},
        pages = {1763-1778},
          doi = {10.1093/mnras/stad1016},
archivePrefix = {arXiv},
       eprint = {2212.11301},
 primaryClass = {astro-ph.GA},
       adsurl = {https://ui.adsabs.harvard.edu/abs/2023MNRAS.522.1763G}
}

@ARTICLE{JermynBauerSchwab_2023ApJS..265...15J,
       author = {{Jermyn}, Adam S. and {Bauer}, Evan B. and {Schwab}, Josiah and {Farmer}, R. and {Ball}, Warrick H. and {Bellinger}, Earl P. and {Dotter}, Aaron and {Joyce}, Meridith and {Marchant}, Pablo and {Mombarg}, Joey S.~G. and et al.},
        title = "{Modules for Experiments in Stellar Astrophysics (MESA): Time-dependent Convection, Energy Conservation, Automatic Differentiation, and Infrastructure}",
      journal = {\apjs},
         year = 2023,
        month = mar,
       volume = {265},
       number = {1},
          eid = {15},
        pages = {15},
          doi = {10.3847/1538-4365/acae8d},
archivePrefix = {arXiv},
       eprint = {2208.03651},
 primaryClass = {astro-ph.SR},
       adsurl = {https://ui.adsabs.harvard.edu/abs/2023ApJS..265...15J}
}

@ARTICLE{HiraiPodsiadlowski_2022MNRAS.517.4544H,
       author = {{Hirai}, Ryosuke and {Podsiadlowski}, Philipp},
        title = "{Neutron stars colliding with binary companions: formation of hypervelocity stars, pulsar planets, bumpy superluminous supernovae and Thorne-{\.Z}ytkow objects}",
      journal = {\mnras},
         year = 2022,
        month = dec,
       volume = {517},
       number = {3},
        pages = {4544-4556},
          doi = {10.1093/mnras/stac3007},
archivePrefix = {arXiv},
       eprint = {2208.00915},
 primaryClass = {astro-ph.HE},
       adsurl = {https://ui.adsabs.harvard.edu/abs/2022MNRAS.517.4544H}
}

@ARTICLE{CantielloJermynLin_2021ApJ...910...94C,
       author = {{Cantiello}, Matteo and {Jermyn}, Adam S. and {Lin}, Douglas N.~C.},
        title = "{Stellar Evolution in AGN Disks}",
      journal = {\apj},
         year = 2021,
        month = apr,
       volume = {910},
       number = {2},
          eid = {94},
        pages = {94},
          doi = {10.3847/1538-4357/abdf4f},
archivePrefix = {arXiv},
       eprint = {2009.03936},
 primaryClass = {astro-ph.SR},
       adsurl = {https://ui.adsabs.harvard.edu/abs/2021ApJ...910...94C}
}

@ARTICLE{Cruz-OsorioRezzolla_2020ApJ...894..147C,
       author = {{Cruz-Osorio}, A. and {Rezzolla}, L.},
        title = "{Common-envelope Dynamics of a Stellar-mass Black Hole: General Relativistic Simulations}",
      journal = {\apj},
         year = 2020,
        month = may,
       volume = {894},
       number = {2},
          eid = {147},
        pages = {147},
          doi = {10.3847/1538-4357/ab89aa},
archivePrefix = {arXiv},
       eprint = {2004.13782},
 primaryClass = {gr-qc},
       adsurl = {https://ui.adsabs.harvard.edu/abs/2020ApJ...894..147C}
}

@ARTICLE{TamayoReinShi_2020MNRAS.491.2885T,
       author = {{Tamayo}, Daniel and {Rein}, Hanno and {Shi}, Pengshuai and {Hernandez}, David M.},
        title = "{REBOUNDx: a library for adding conservative and dissipative forces to otherwise symplectic N-body integrations}",
      journal = {\mnras},
         year = 2020,
        month = jan,
       volume = {491},
       number = {2},
        pages = {2885-2901},
          doi = {10.1093/mnras/stz2870},
archivePrefix = {arXiv},
       eprint = {1908.05634},
 primaryClass = {astro-ph.EP},
       adsurl = {https://ui.adsabs.harvard.edu/abs/2020MNRAS.491.2885T}
}

@ARTICLE{PaxtonSmolecSchwab_2019ApJS..243...10P,
       author = {{Paxton}, Bill and {Smolec}, R. and {Schwab}, Josiah and {Gautschy}, A. and {Bildsten}, Lars and {Cantiello}, Matteo and {Dotter}, Aaron and {Farmer}, R. and {Goldberg}, Jared A. and {Jermyn}, Adam S. and et al.},
        title = "{Modules for Experiments in Stellar Astrophysics (MESA): Pulsating Variable Stars, Rotation, Convective Boundaries, and Energy Conservation}",
      journal = {\apjs},
         year = 2019,
        month = jul,
       volume = {243},
       number = {1},
          eid = {10},
        pages = {10},
          doi = {10.3847/1538-4365/ab2241},
archivePrefix = {arXiv},
       eprint = {1903.01426},
 primaryClass = {astro-ph.SR},
       adsurl = {https://ui.adsabs.harvard.edu/abs/2019ApJS..243...10P}
}

@ARTICLE{PaxtonSchwabBauer_2018ApJS..234...34P,
       author = {{Paxton}, Bill and {Schwab}, Josiah and {Bauer}, Evan B. and {Bildsten}, Lars and {Blinnikov}, Sergei and {Duffell}, Paul and {Farmer}, R. and {Goldberg}, Jared A. and {Marchant}, Pablo and {Sorokina}, Elena and et al.},
        title = "{Modules for Experiments in Stellar Astrophysics (MESA): Convective Boundaries, Element Diffusion, and Massive Star Explosions}",
      journal = {\apjs},
         year = 2018,
        month = feb,
       volume = {234},
       number = {2},
          eid = {34},
        pages = {34},
          doi = {10.3847/1538-4365/aaa5a8},
archivePrefix = {arXiv},
       eprint = {1710.08424},
 primaryClass = {astro-ph.SR},
       adsurl = {https://ui.adsabs.harvard.edu/abs/2018ApJS..234...34P}
}

@ARTICLE{Murguia-BerthierMacLeodRamirez-Ruiz_2017ApJ...845..173M,
       author = {{Murguia-Berthier}, Ariadna and {MacLeod}, Morgan and {Ramirez-Ruiz}, Enrico and {Antoni}, Andrea and {Macias}, Phillip},
        title = "{Accretion Disk Assembly During Common Envelope Evolution: Implications for Feedback and LIGO Binary Black Hole Formation}",
      journal = {\apj},
         year = 2017,
        month = aug,
       volume = {845},
       number = {2},
          eid = {173},
        pages = {173},
          doi = {10.3847/1538-4357/aa8140},
archivePrefix = {arXiv},
       eprint = {1705.04698},
 primaryClass = {astro-ph.SR},
       adsurl = {https://ui.adsabs.harvard.edu/abs/2017ApJ...845..173M}
}

@ARTICLE{OhlmannRopkePakmor_2017A&A...599A...5O,
       author = {{Ohlmann}, Sebastian T. and {R{\"o}pke}, Friedrich K. and {Pakmor}, R{\"u}diger and {Springel}, Volker},
        title = "{Constructing stable 3D hydrodynamical models of giant stars}",
      journal = {\aap},
         year = 2017,
        month = mar,
       volume = {599},
          eid = {A5},
        pages = {A5},
          doi = {10.1051/0004-6361/201629692},
archivePrefix = {arXiv},
       eprint = {1612.00008},
 primaryClass = {astro-ph.SR},
       adsurl = {https://ui.adsabs.harvard.edu/abs/2017A&A...599A...5O}
}

@ARTICLE{PaxtonMarchantSchwab_2015ApJS..220...15P,
       author = {{Paxton}, Bill and {Marchant}, Pablo and {Schwab}, Josiah and {Bauer}, Evan B. and {Bildsten}, Lars and {Cantiello}, Matteo and {Dessart}, Luc and {Farmer}, R. and {Hu}, H. and {Langer}, N. and et al.},
        title = "{Modules for Experiments in Stellar Astrophysics (MESA): Binaries, Pulsations, and Explosions}",
      journal = {\apjs},
         year = 2015,
        month = sep,
       volume = {220},
       number = {1},
          eid = {15},
        pages = {15},
          doi = {10.1088/0067-0049/220/1/15},
archivePrefix = {arXiv},
       eprint = {1506.03146},
 primaryClass = {astro-ph.SR},
       adsurl = {https://ui.adsabs.harvard.edu/abs/2015ApJS..220...15P}
}

@ARTICLE{MacLeodRamirez-Ruiz_2015ApJ...803...41M,
       author = {{MacLeod}, Morgan and {Ramirez-Ruiz}, Enrico},
        title = "{Asymmetric Accretion Flows within a Common Envelope}",
      journal = {\apj},
         year = 2015,
        month = apr,
       volume = {803},
       number = {1},
          eid = {41},
        pages = {41},
          doi = {10.1088/0004-637X/803/1/41},
archivePrefix = {arXiv},
       eprint = {1410.3823},
 primaryClass = {astro-ph.SR},
       adsurl = {https://ui.adsabs.harvard.edu/abs/2015ApJ...803...41M}
}

@ARTICLE{ReinSpiegel_2015MNRAS.446.1424R,
       author = {{Rein}, Hanno and {Spiegel}, David S.},
        title = "{IAS15: a fast, adaptive, high-order integrator for gravitational dynamics, accurate to machine precision over a billion orbits}",
      journal = {\mnras},
         year = 2015,
        month = jan,
       volume = {446},
       number = {2},
        pages = {1424-1437},
          doi = {10.1093/mnras/stu2164},
archivePrefix = {arXiv},
       eprint = {1409.4779},
 primaryClass = {astro-ph.EP},
       adsurl = {https://ui.adsabs.harvard.edu/abs/2015MNRAS.446.1424R}
}

@ARTICLE{MacLeodRamirez-Ruiz_2015ApJ...798L..19M,
       author = {{MacLeod}, Morgan and {Ramirez-Ruiz}, Enrico},
        title = "{On the Accretion-fed Growth of Neutron Stars during Common Envelope}",
      journal = {\apjl},
         year = 2015,
        month = jan,
       volume = {798},
       number = {1},
          eid = {L19},
        pages = {L19},
          doi = {10.1088/2041-8205/798/1/L19},
archivePrefix = {arXiv},
       eprint = {1410.5421},
 primaryClass = {astro-ph.SR},
       adsurl = {https://ui.adsabs.harvard.edu/abs/2015ApJ...798L..19M}
}

@ARTICLE{LevesqueMasseyZytkow_2014MNRAS.443L..94L,
       author = {{Levesque}, E.~M. and {Massey}, P. and {Zytkow}, A.~N. and {Morrell}, N.},
        title = "{Discovery of a Thorne-Zytkow object candidate in the Small Magellanic Cloud.}",
      journal = {\mnras},
         year = 2014,
        month = sep,
       volume = {443},
        pages = {L94-L98},
          doi = {10.1093/mnrasl/slu080},
archivePrefix = {arXiv},
       eprint = {1406.0001},
 primaryClass = {astro-ph.SR},
       adsurl = {https://ui.adsabs.harvard.edu/abs/2014MNRAS.443L..94L}
}

@ARTICLE{PaxtonCantielloArras_2013ApJS..208....4P,
       author = {{Paxton}, Bill and {Cantiello}, Matteo and {Arras}, Phil and {Bildsten}, Lars and {Brown}, Edward F. and {Dotter}, Aaron and {Mankovich}, Christopher and {Montgomery}, M.~H. and {Stello}, Dennis and {Timmes}, F.~X. and et al.},
        title = "{Modules for Experiments in Stellar Astrophysics (MESA): Planets, Oscillations, Rotation, and Massive Stars}",
      journal = {\apjs},
         year = 2013,
        month = sep,
       volume = {208},
       number = {1},
          eid = {4},
        pages = {4},
          doi = {10.1088/0067-0049/208/1/4},
archivePrefix = {arXiv},
       eprint = {1301.0319},
 primaryClass = {astro-ph.SR},
       adsurl = {https://ui.adsabs.harvard.edu/abs/2013ApJS..208....4P}
}

@ARTICLE{IvanovaJusthamChen_2013A&ARv..21...59I,
       author = {{Ivanova}, N. and {Justham}, S. and {Chen}, X. and {De Marco}, O. and {Fryer}, C.~L. and {Gaburov}, E. and {Ge}, H. and {Glebbeek}, E. and {Han}, Z. and {Li}, X.-D. and et al.},
        title = "{Common envelope evolution: where we stand and how we can move forward}",
      journal = {\aapr},
         year = 2013,
        month = feb,
       volume = {21},
          eid = {59},
        pages = {59},
          doi = {10.1007/s00159-013-0059-2},
archivePrefix = {arXiv},
       eprint = {1209.4302},
 primaryClass = {astro-ph.HE},
       adsurl = {https://ui.adsabs.harvard.edu/abs/2013A&ARv..21...59I}
}

@ARTICLE{RickerTaam_2012ApJ...746...74R,
       author = {{Ricker}, Paul M. and {Taam}, Ronald E.},
        title = "{An AMR Study of the Common-envelope Phase of Binary Evolution}",
      journal = {\apj},
         year = 2012,
        month = feb,
       volume = {746},
       number = {1},
          eid = {74},
        pages = {74},
          doi = {10.1088/0004-637X/746/1/74},
archivePrefix = {arXiv},
       eprint = {1107.3889},
 primaryClass = {astro-ph.SR},
       adsurl = {https://ui.adsabs.harvard.edu/abs/2012ApJ...746...74R}
}

@ARTICLE{ReinLiu_2012A&A...537A.128R,
       author = {{Rein}, H. and {Liu}, S.-F.},
        title = "{REBOUND: an open-source multi-purpose N-body code for collisional dynamics}",
      journal = {\aap},
         year = 2012,
        month = jan,
       volume = {537},
          eid = {A128},
        pages = {A128},
          doi = {10.1051/0004-6361/201118085},
archivePrefix = {arXiv},
       eprint = {1110.4876},
 primaryClass = {astro-ph.EP},
       adsurl = {https://ui.adsabs.harvard.edu/abs/2012A&A...537A.128R}
}

@ARTICLE{BallToutZytkow_2011MNRAS.414.2751B,
       author = {{Ball}, Warrick H. and {Tout}, Christopher A. and {{\.Z}ytkow}, Anna N. and {Eldridge}, John J.},
        title = "{The structure and evolution of quasi-stars}",
      journal = {\mnras},
         year = 2011,
        month = jul,
       volume = {414},
       number = {3},
        pages = {2751-2762},
          doi = {10.1111/j.1365-2966.2011.18591.x},
archivePrefix = {arXiv},
       eprint = {1102.5098},
 primaryClass = {astro-ph.HE},
       adsurl = {https://ui.adsabs.harvard.edu/abs/2011MNRAS.414.2751B}
}

@ARTICLE{PaxtonBildstenDotter_2011ApJS..192....3P,
       author = {{Paxton}, Bill and {Bildsten}, Lars and {Dotter}, Aaron and {Herwig}, Falk and {Lesaffre}, Pierre and {Timmes}, Frank},
        title = "{Modules for Experiments in Stellar Astrophysics (MESA)}",
      journal = {\apjs},
         year = 2011,
        month = jan,
       volume = {192},
       number = {1},
          eid = {3},
        pages = {3},
          doi = {10.1088/0067-0049/192/1/3},
archivePrefix = {arXiv},
       eprint = {1009.1622},
 primaryClass = {astro-ph.SR},
       adsurl = {https://ui.adsabs.harvard.edu/abs/2011ApJS..192....3P}
}

@BOOK{Chandrasekhar_1957isss.book.....C,
       author = {{Chandrasekhar}, Subrahmanyan},
        title = "{An introduction to the study of stellar structure.}",
         year = 1957,
       adsurl = {https://ui.adsabs.harvard.edu/abs/1957isss.book.....C}
}
\bibliographystyle{aasjournal}



\end{document}